\documentclass[sigconf, nonacm, pdfa]{acmart}

\newcommand\vldbdoi{XX.XX/XXX.XX}

\newcommand\vldbavailabilityurl{https://github.com/NLGithubWP/indices}

\usepackage{mathrsfs}
\usepackage{booktabs}
\usepackage{multirow}
\usepackage{algorithm}
\usepackage{algpseudocode}
\usepackage{subcaption}
\usepackage{graphicx}
\usepackage{xspace}
\usepackage{float}
\usepackage[most]{tcolorbox}
\usepackage{balance}
\usepackage{makecell}

\usepackage{pifont}

\usepackage{listings}
\usepackage{enumitem}
\setlist[itemize]{align=parleft,left=0pt..1em}

\lstdefinestyle{SQL}{
    language=SQL,
    basicstyle=\ttfamily,
    keywordstyle=\color{blue}\bfseries,
    commentstyle=\color{green!40!black},
    stringstyle=\color{red},
    showstringspaces=false,
    breaklines=true,
    xleftmargin=\parindent, 
    numbers=none, 
    captionpos=b, 
    columns=fullflexible 
}

\newcommand{\Name}{LEADS\xspace}

\newcommand{\technique}{SQL-aware dynamic model slicing\xspace}
\newcommand{\Technique}{SQL-Aware Dynamic Model Slicing\xspace}

\newcommand{\red}[1]{{\color{black}#1}}

\newcommand{\bluerevise}[1]{{\color{black}#1}}
\newcommand{\revision}[1]{{\color{black}#1}}

\newcommand{\term}[1]{{\color{black}#1}}

\newcommand{\ignore}[1]{}

\newcommand{\highlight}[1]{\textbf{#1}}

\usepackage[a-2b]{pdfx}

\newtcolorbox{myquote}[1][]{
    colback=black!10,
    colframe=black!10,
    notitle,
    sharp corners,
    borderline west={2pt}{0pt}{red!80!black},
    enhanced,
    breakable,
}

\begin{document}

\title[Powering In-Database Dynamic Model Slicing for Structured Data Analytics]{Powering In-Database Dynamic Model Slicing \\for Structured Data Analytics}


\author{Lingze Zeng}
\affiliation{%
  \institution{National University of Singapore}
}
\email{lingze@comp.nus.edu.sg}

\author{Naili Xing}
\affiliation{%
  \institution{National University of Singapore}
}
\email{xingnl@comp.nus.edu.sg}

\author{Shaofeng Cai}
\affiliation{%
  \institution{National University of Singapore}
  \country{}
}
\email{shaofeng@comp.nus.edu.sg}

\author{Gang Chen}
\affiliation{%
  \institution{Zhejiang University}
}
\email{cg@zju.edu.cn}

\author{Beng Chin Ooi}
\affiliation{%
  \institution{National University of Singapore}
  \country{}
}
\email{ooibc@comp.nus.edu.sg}

\author{Jian Pei}
\affiliation{%
  \institution{Duke University}
}
\email{j.pei@duke.edu}

\author{Yuncheng Wu}
\affiliation{%
  \institution{Renmin University of China}
}
\email{wuyuncheng@ruc.edu.cn}


\begin{abstract}
\label{sec:abstract}
Relational database management systems (RDBMS) are widely used for the storage of structured data.
To derive insights beyond statistical aggregation, 
we typically have to extract specific subdatasets from the database using conventional database operations, and then apply deep neural networks (DNN) training and inference on these subdatasets in a separate analytics system.
The process can be prohibitively expensive,
especially when there are various subdatasets extracted for different analytical purposes.
This calls for efficient in-database support of advanced analytical methods.

In this paper, we introduce \Name, a novel \technique 
technique to customize models for specified SQL queries.
\Name improves the predictive modeling of structured data via the mixture of experts (MoE) and maintains efficiency by a SQL-aware gating network.
At the core of \Name is the construction of a general model with multiple expert sub-models trained over the database.
The MoE scales up the modeling capacity, enhances effectiveness, and preserves efficiency by activating necessary experts via the SQL-aware gating network during inference.
\red{
To support in-database analytics,
we build an inference extension 
that integrates \Name onto PostgreSQL.
}
Our extensive experiments on real-world datasets demonstrate that \Name~ consistently outperforms the baseline models, and the in-database inference extension delivers a considerable reduction in inference latency compared to traditional solutions.

\end{abstract}


\maketitle


\section{Introduction}
\label{sec:introduction}

Relational Database Management Systems (RDBMS)
are extensively employed as the primary storage solution for structured data across various applications \cite{khamis2020learning,nikolic2020f,li2019enabling, rendle2013scaling}.
They serve as a fundamental infrastructure for various domains and are critical to the operation of numerous businesses \cite{jia2023robust, zheng2020tracer,he2017neural}.
In the contemporary business landscape,
\term{structured data analytics} via databases has become an indispensable component for driving business growth and success \cite{jia2023robust, zheng2020tracer,he2017neural,ParkSBSIK22}.
Traditional structured data analytics approaches rely on database-driven filtering or aggregation operations to derive insights.
However, these insights only offer a limited statistical view, which often fails to capture the complexity and intricacies of the underlying patterns \cite{gray1997data,rajakumari2014efficient}.
Fortunately, recent advancements in Deep Neural Networks (DNNs) open up new horizons for advanced analytics beyond simple statistical aggregation \cite{gardner1998artificial, lian2018xdeepfm, cheng2020adaptive, cai2021arm}.

At its core, exploiting DNNs for advanced structured data analytics comprises two main \term{phases}: training and inference \cite{gardner1998artificial}.
The former primarily involves the \term{construction} of a DNN model and the training of this model on targeted data, while the latter utilizes the trained model to make predictions on new data.
Notably, to deliver advanced DNN-driven analytics for informed decision-making, effectiveness, and efficiency are the two most important \term{metrics} to optimize for \cite{davis2006relationship,kwon2019understanding,cai2019modelvldb}.
Specifically, effectiveness focuses on the inference phase, measuring the extent to which the predictions delivered by the model are accurate.
Meanwhile, efficiency evaluates the requirements of the model in terms of \term{\textit{response time} and \textit{computational resources}} in both phases \cite{kwon2019understanding}.

In real-world scenarios, \term{analysts} are often more interested in performing analytics on
specific subsets of data. For instance, they may assess trends among patients diagnosed with a particular disease, or study behaviors of consumers in a certain age group.
\bluerevise{
Consider the scenario illustrated in Figure~\ref{fig:motivation}, where an analyst examines the income situation within a specific demographic. This analysis informs critical decision-making, such as setting personal loan limits and adjusting local commodity prices.}
Naturally, the analyst seeks to \term{build} an effective predictive model, delivering accurate predictions for these subsets of tuples, meanwhile executing predictions efficiently with minimal response time and computational resources.
However, there are two main challenges in achieving this objective.

First, achieving efficient training for effective predictive modeling across analyst-specified subdatasets is challenging.
Conventionally, a single \term{general model} is trained to support inference across all data tuples \ignore{\red{[refs]}}\cite{he2017neural,khamis2020learning,gardner1998artificial}.
This approach is efficient and requires training only one model.
However, such a model, optimized to capture the common patterns and general behaviors of the \term{whole dataset}, is likely not as effective in providing accurate predictions as a dedicated model trained on a specific subdataset of interest.
Taking the scenario in Figure~\ref{fig:motivation} as an example, a model trained explicitly for the group of men living in NYC would probably identify finer-grained patterns and behaviors pertinent to this subdataset. Given sufficient training instances, this dedicated model could outperform the general model significantly.
Nonetheless, training a separate model for each subdataset is computationally prohibitive due to the combinatorial nature of potential subdatasets.


Secondly, enhancing the efficiency of inference execution for subdatasets queried from the RDBMS is also challenging from a system perspective.
A major challenge is the coordination of the RDBMS data processing and the execution of inference tasks.
Many existing solutions support the inference process using two separate systems~\cite{rafiki,clipper,zhang2021distributed}, which requires transferring the inference data from an RDBMS to the inference system.
However, this process is time-consuming, susceptible to errors, and may violate privacy and security requirements \cite{xu2022database}.
More critically, additional data transfer overhead is introduced to adversely affect inference efficiency.

To address the above challenges, we propose
a novel SQ\textbf{\underline{L}}-awar\textbf{\underline{E}} dyn\textbf{\underline{A}}mic mo\textbf{\underline{D}}el \textbf{\underline{S}}licing (\Name) technique, which 
makes use of the meta-information in SQL queries to dynamically customize a predictive model, deriving meaningful insights in analysis tasks. Specifically, \Name first constructs a high-capacity general model consisting of multiple replicas of the base model.
These replicas, termed as \textit{experts} in MoE, are trained to specialize in different problem subspaces for effective predictive modeling.
To enhance effectiveness via MoE without incurring reduced inference efficiency, LEADS incorporates a SQL-aware gating network, which generates sparse \textit{gating weights} based on the SQL query embedding to select necessary experts.
Such a \textit{sliced model} is trained 
to tailor for SQL queries and dedicated to the specified subdataset for inference effectiveness while maintaining efficiency.
To further improve efficiency, we build an extension to integrate the inference process within the database via User-Defined-Function (UDF) with three optimizations:
efficient execution allocation, memory sharing, and state caching. 
These approaches obviate the data transfer in separate systems, reduce data copying overhead, and reduce the cost associated with frequent model loading.
We summarize our main contributions as follows.
\begin{itemize}
    \item We formulate the \term{SQL-aware structured data analytics} problem, which requires efficient and effective predictive modeling on subdatasets specified by SQL queries.
    To the best of our knowledge, this is the first work for solving the problem.

    \item We propose a novel \technique technique \Name, which scales up the modeling capacity by replicating the base model as experts and devises a SQL-aware gating network to dynamically customize models for analytics tasks on SQL-specified subdatasets.
    
    \item We implement an in-database inference extension on PostgreSQL to support SQL-aware structured data analytics, which 
    incorporates three optimizations to improve inference efficiency.
    
    \item We conduct extensive experiments on five real-world datasets, which confirms the effectiveness of \Name, with up to $3.95\%$ improvement in AUC for given data workloads, \red{while the inference extension achieves up to $2.06$x speedup in terms of efficiency compared with the baseline approaches.}
\end{itemize}
    \red{
  We  have integrated \Name with the extension onto NeurDB  \cite{ooi2024neurdb,zhao2024neurdbcidr},
  our ongoing AI-powered autonomous data system implementation. 
}

\begin{figure}
\centering
\setcounter{figure}{0}
\includegraphics[width=0.42\textwidth]{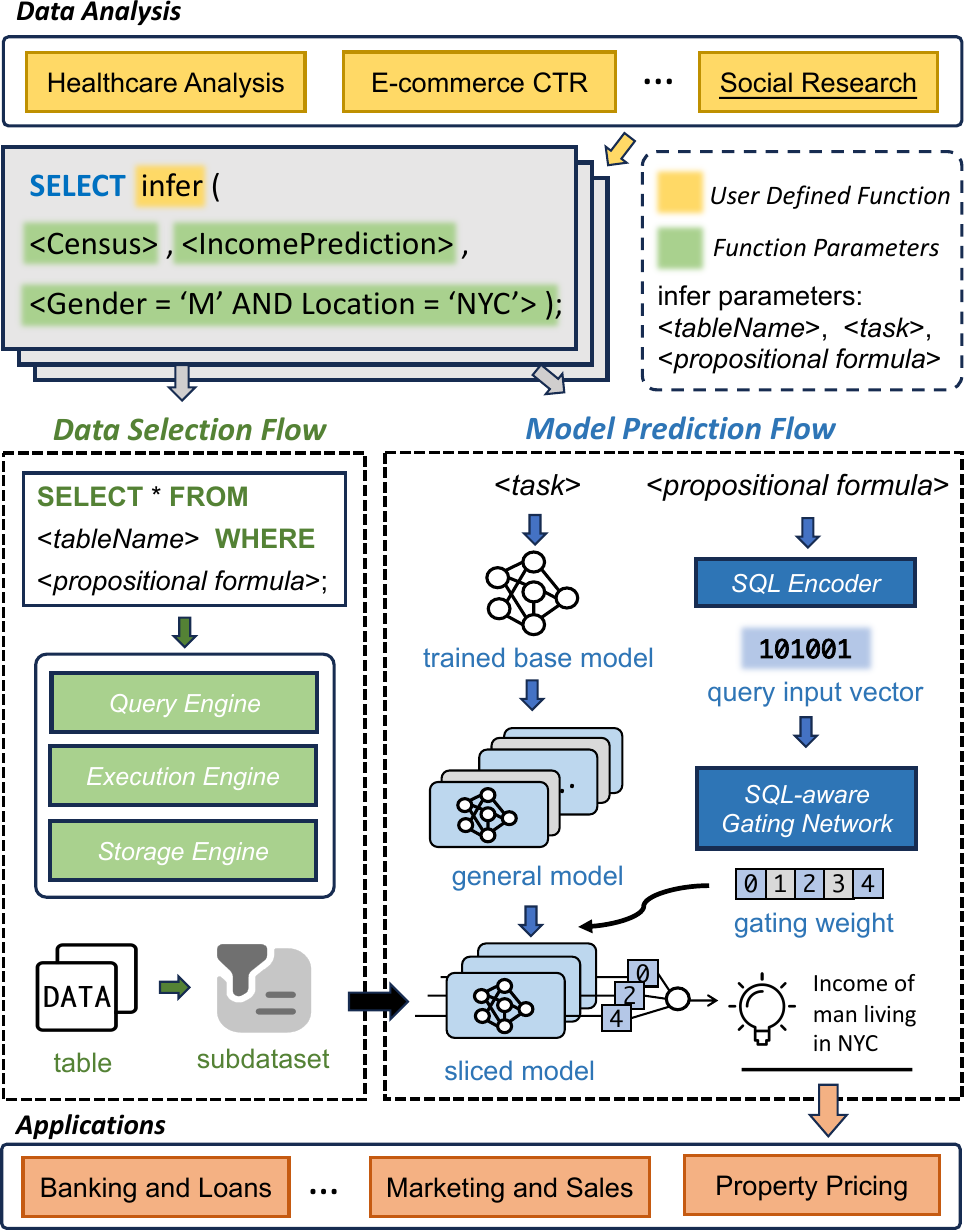}
\caption{\bluerevise{A workflow of supporting in-database analytics on income using SQL-aware dynamic model slicing.}}
\label{fig:motivation}
\end{figure}
\red{
The paper is structured as follows.
We introduce preliminaries in Section~\ref{sec:preliminaries}.
We present details of \Name in Section~\ref{sec:sql_aware_network}
and discuss the implementation in Section~\ref{sec:indb_inference}.
Experimental results are presented in Section~\ref{sec:experiment}. 
We review related work in Section~\ref{sec:relatedWork} and conclude  in Section~\ref{sec:conclusion}.
}

\section{Preliminaries}
\label{sec:preliminaries}
\bluerevise{In this section, we introduce the preliminaries of predictive modeling on structured data} and present two key techniques central to our system, namely Mixture of Experts (MoE) for 
scaling up the model capacity 
while maintaining its inference efficiency via conditional computation \cite{shazeer2017outrageously},
and sparsemax \cite{martins2016softmax} for the informed selection of active experts to enhance efficiency.
\bluerevise{We also formally define the research problem, referred to as \texttt{SQL-aware predictive modeling.}}
Scalars, vectors, and matrices in the following part are denoted by $x$, $\mathbf{x}$ and $\mathbf{X}$ respectively.

\vspace{1mm}
\bluerevise{
\noindent \textbf{Structured Data} 
(or relational data, tabular data) ~\cite{SD_arasu2003extracting,SD_bakir2007predicting} refers to the type of data that can be represented as tables.
It is typically stored in a series of tables (relations) $\{\mathbf{T_1}, \mathbf{T_2},...\}$ of columns and rows, which can be retrieved from a relational database with SQL operations, e.g., the projection, join, and aggregation. Tables in structured data are interconnected via \textit{foreign key} attributes, where the value of a column in one table points to a unique row in another table.
Technically, the structured data can be viewed as a logical table $\textbf{T}$ comprising $N$ rows and $M$ columns.
Each row is a tuple $ {\mathbf{x}} = (x_1,x_2,...,x_M)$, serving as a feature vector, where $x_i$ is the value of the $i$-th attribute and can be either numerical or categorical.
}

\vspace{1mm}
\noindent \textbf{Mixture of Experts} (MoE)
~\cite{jordan1994hierarchical,waterhouse1995bayesian,shazeer2017outrageously} is a general ensemble learning and conditional computation technique to scale up the modeling capacity without incurring much computational overhead. 
\bluerevise{MoE is particularly effective when the data exhibits complex patterns or variations~\cite{riquelme2021scaling, fedus2022switch, zhou2022mixture}.
There are two main components in an MoE layer: expert models and a gating network. Expert models can be composed of homogeneous models that share the same architecture. They are adopted to divide problem space into different regions, where each expert specializes in handling a certain sub-region.
The gating network outputs a set of gating weights for a given input, dynamically determining weights assigned to respective experts.}

Denoting the gating weights and outputs of experts as $\mathbf{w} = [w_1, w_2, ..., w_K]$ and $\mathbf{H} = [\mathbf{h}_1, \mathbf{h}_2 ..., \mathbf{h}_K]$ respectively, where $K$ is the number of experts and $\mathbf{h}_i$ is the output of the $i$-th expert. the output of the MoE is a weighted average of these experts: $\hat{\mathbf{y}} = \sum_{i = 1}^K w_i \mathbf{h}_i$.
During training, the MoE model optimizes the gating network and experts simultaneously.
The gating network learns to assign appropriate weight to experts, while the experts learn to make accurate predictions within their respective regions of expertise. 

MoE has found extensive application in various domains, notably in the large language model GPT-4~\cite{gpt4} for texts and the large vision-language model MoE-LLaVA~\cite{llava} for images, which combines the benefits of large model capacity with efficient computation, by only engaging a fraction of the model parameters for each input.
In \Name, we apply the MoE technique to structured data, intending to harness its scalable modeling capacity for enhanced predictive effectiveness and efficiency.

\vspace{1mm}
\noindent \textbf{Sparsemax} \bluerevise{\cite{martins2016softmax, entmax, cai2021arm}}.
Softmax transformation is a crucial function in the gating network, which maps an input vector $\mathbf{z}$ into a probability distribution $\mathbf{p}$ whose probabilities correspond proportionally to the exponential of its input values, i.e., 
$\mathrm {softmax(z_j)} = \frac{\mathrm{exp(z_j)}}{\mathrm{\sum_{i}exp(z_i)}}$.
The output can be denoted as the class probabilities or weights indicating the importance of corresponding inputs.

The softmax function is extensively used in DNNs due to its differentiable and convex properties.
However, the output probabilities of the softmax function are dense, leading to less interpretability and effectiveness~\cite{goodfellow20166,cai2021arm}.
To overcome this limitation, $\alpha$-entmax is proposed to generalize the dense and sparse softmax, offering a parameter to adjust the sparsity level of probability distribution.
Specifically, given a d-dimension probability as 
$\rm  \Delta ^d := \{\mathbf{p} \in \mathbb{R}^d : \mathbf{p} \geq 0. \text{ } ||\mathbf{p}||_1 = 1\}$, $\alpha$-entmax~\cite{entmax} can be interpreted in the variational form with Tsallis $\alpha$-entropies $\rm H_{\alpha}(\mathbf{p})$~\cite{tsallis1988possible}:

\begin{equation}
\rm
\alpha\text{-}entmax(\mathbf{z}) = \mathop{argmax}\limits_{\mathbf{p} \in \Delta^d}{\mathbf{p}^T\mathbf{z} + H_{\alpha}(\mathbf{p})}
\end{equation}

\noindent
where $\mathrm{H_{\alpha}(\mathbf{p})} = - \frac{1}{\alpha(\alpha -1)} \sum_j(p_j - p^{\alpha}_j)$ if $\rm \alpha \neq 1$, else $\rm H_1(\mathbf{p}) =  -\sum_jp_j \log p_j$, the Shannon entropy.
A larger $\alpha$ causes $\alpha$-entmax to assign more zero in the probability distribution.
\ignore{
Another appealing property is that the hyper-parameter $\alpha$, which controls the shape and sparsity of the mapping, can be learned \term{adaptively} to the predictive task in the training stage.
Let $\mathbf{p}^{*} = \alpha\text{-entmax(\textbf{z})}$ denote the distribution $\widetilde{p_i} = (p^{*}_i)^{2-\alpha} / \sum_j (p^{*}_j)^{2-\alpha}$ and the Shannon entropy $h_i = -(p^{*}_i)log(p^{*}_i)$.
The gradient of $\alpha$ is derived as:
$\frac{\partial \ {\alpha\text{-entmax(\textbf{z})}}}{\partial \alpha} =  \frac{(p^{*}_i) - \widetilde{p_i}}{{(\alpha - 1)}^2} + \frac{h_i -\widetilde{p_i} \sum_j h_j}{\alpha - 1}, \ \alpha > 1$
, which can be optimized end-to-end together with the parameters of the predictive model~\cite{entmax}.}\bluerevise{Sparsemax is adopted in \Name to sparsify the activated experts in MoE to enhance predictive efficiency.}

\vspace{1.5mm}
\noindent \textbf{Problem Formulation}.
In structured data analytics, data analysts typically focus on specific subdataset characterized by shared attributes.
For example, analysts may assess the readmission rates among patients diagnosed with a certain disease,
or predict the e-commerce click-through rate (CTR) within a particular age group.
Typically, for advanced analytical tasks that involve prediction, \texttt{WHERE} statement in a SQL query is executed first to select relevant tuples, to which DNNs are applied subsequently for inference.
We refer to this process as \texttt{SQL-aware predictive modeling}. 

Given structured data $\mathbf{T}$ as illustrated in Section~\ref{sec:preliminaries} and a SQL query denoted by $q$, there are two main steps in SQL-aware predictive modeling: data selection and model prediction.
Corresponding to relational algebra, a generalized SQL query selection $q$ is expressed as $\sigma_{\varphi}(\textbf{T})$, where $\sigma$ is the unary operator for selection and $\varphi$ is the propositional formula in $q$.
Typically, $\varphi$ consists of multiple predicates connected by logical operators. 
The selection $\sigma_{\varphi}(\textbf{T})$ retrieves all tuples in table $\textbf{T}$ that satisfies $\varphi$, formally defined as  $\sigma_\varphi (\textbf{T}) = \{ \textbf{x}: \textbf{x} \in \textbf{T}, \varphi (\textbf{x})\}$.
For simplicity, the subdataset retrieved by the SQL query $q$ is denoted as $\textbf{T}_\varphi = \{\textbf{x}_1, \textbf{x}_2, \cdots, \textbf{x}_{n}\}$, where $n$ is the number of tuples.
Each tuple $\textbf{x}_i \in \mathbb{R}^M$ 
\ignore{in $\textbf{T}_\varphi$} comprises $M$ attributes,
and $\textbf{x}_i$ can be represented as a vector of features, i.e., $ \textbf{x}_i = [ x_{i,1},x_{i,2},\cdots, x_{i, M} ] $.
DNNs are then applied to perform prediction on these selected tuples, e.g., to predict the labels $\textbf{y} = \{y_1, y_2,\cdots, y_n\}$, aiming to derive meaningful insights, such as patients readmission rates in healthcare analytics or CTR in e-commerce.
Technically, \texttt{SQL-aware predictive modeling} refers to making predictions on a selected subset of tuples retrieved from a logical table $\textbf{T}$ based on a SQL query $q$ with a propositional formula $\varphi$.

\ignore{
\vspace{1mm}
\noindent
\red{
\textbf{Assumptions.} 
Our system operates under the assumption that the database remains static, where modifications to the schema or variations in data patterns require fine-tuning or retraining.
The current system supports SQL queries featuring equality selection predicates, a common practice in real-world applications.
Notably, the system can be extended to support Select-Project-Join (SPJ) queries, where results from joining multiple interconnected tables need to be viewed as one single logical table.
With this consolidated table view, \Name can train a predictive model and enable SQL queries that perform selections on it to utilize the trained model for supporting SPJ queries.
}
}

\section{\Technique}
\label{sec:sql_aware_network}
\bluerevise{In this section, we introduce our SQL-aware dynamic model slicing technique, \Name, for supporting in-database predictive analytics within existing RDBMSs such as PostgreSQL.
Unlike conventional machine learning approaches that make predictions based solely on features of individual instances, \Name identifies common attributes of subdataset as specified by SQL queries and exploits this meta-information for customized, efficient and effective predictive modeling.
For example, as illustrated in Figure~\ref{fig:SQLEncoding}, the SQL query specifies instances sharing attributes of gender (male), age (24), and location (NYC).
\Name leverages these propositional formulas, i.e., $\varphi$ in SQL queries, to customize the predictive model, thereby enhancing prediction accuracy for the specified subdataset.
}

\bluerevise{We first propose a SQL query encoder to extract the information encapsulated in $\varphi$ into a representation vector for subsequent predictive modeling.
Next, we elaborate on the model architecture and data processing flow of \Name for SQL-aware structured data modeling.
Finally, we introduce the optimization schemes, with two novel regularization terms proposed to balance the effectiveness and efficiency of the modeling process.}

\subsection{SQL Query Encoder}
\label{sec:encoder}
\begin{figure}[t]
  \centering
  \begin{subfigure}[b]{\columnwidth}
    \centering
    \includegraphics[width=0.99\columnwidth]{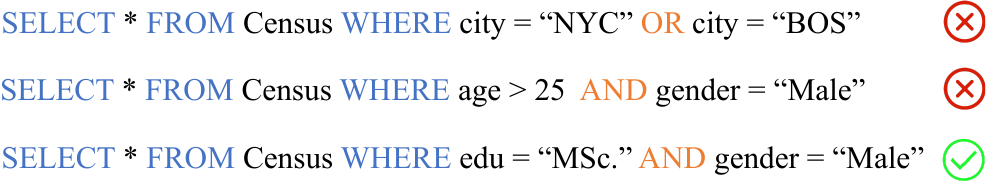}
    \caption{Examples of a primitive SQL query.}
    \label{fig:UnitSQL}
  \end{subfigure}
  
  \vspace{\baselineskip}

  \begin{subfigure}[b]{\columnwidth}
    \centering
    \includegraphics[width=0.99\columnwidth]{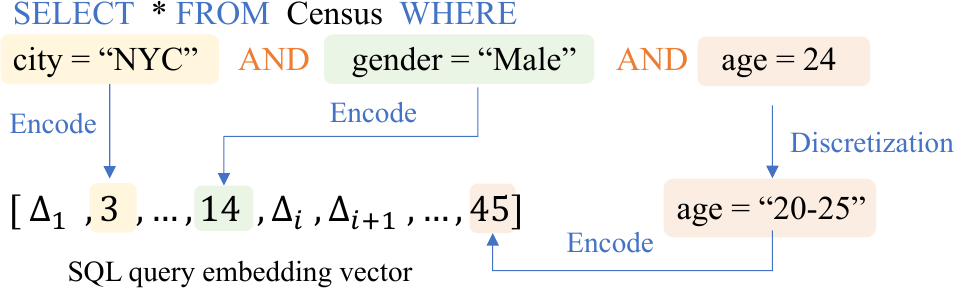}
    \caption{Process of encoding a SQL query.}
    \label{fig:SQLEncoding}
  \end{subfigure}
  \caption{SQL query encoder.}
  \label{fig:SQL}
\end{figure}

In SQL-aware predictive modeling, the WHERE clause filters tuples based on a propositional formula $\varphi$. This formula consists of predicates, each imposing a condition on a specific attribute, like "gender = ‘M’". These predicates are combined using logical operators such as "\texttt{AND}" or "\texttt{OR}" to form a propositional formula. 
\bluerevise{
Given the exponential number of possible predicate combinations, devising a general approach to effectively transform SQL queries into feature vectors for subsequent predictive modeling is challenging.
To achieve this, we focus on individual queries, referred to as \texttt{primitive SQL query}.}
Considering a table $\textbf{T}$ with $M$ attributes, the $j$-th attribute denoted as $A_j$, each attribute is linked to either a numerical or categorical feature. Particularly, each numerical feature needs to be converted into a corresponding categorical feature through discretization, which will be detailed subsequently.
In a primitive SQL query, each attribute $A_j$ may be associated with zero or one predicate, with predicates across attributes conjoined using the logical operator $\land$ (\texttt{AND}).
Technically, a predicate for the attribute $A_j$ in a primitive SQL query can be expressed as $P_j: A_j = a_j$, where $a_j \in \mathcal{D}_j \cup \{ \Delta_j \}$, $\mathcal{D}_j$ represents the domain of possible values for attribute $A_j$, and $\Delta_j$ denotes a default value assigned to $A_j$ when it is not specified in the query.
Figure \ref{fig:UnitSQL} illustrates a valid primitive SQL query example, contrasting with two non-examples.
Thus, the propositional formula $\varphi$ can be represented as:
$$
\varphi = P_1 \land P_2 \land \cdots \land P_M.
$$

\noindent
The objective of the SQL query encoder is to generate a categorical feature vector $\mathbf{q}$ for each primitive SQL query to represent the meta-information from $\varphi$, which is achieved by concatenating the attribute values of the predicates.
Formally, the feature vector of the SQL query encoding can be obtained by:
$$
\mathbf{q} = [q_1, q_2, \cdots, q_M]
$$
\noindent
where $q_j$ is the categorical attribute value for predicate $P_j$.
Figure~\ref{fig:SQLEncoding} demonstrates the transformation of a primitive SQL query into a feature vector.
Notably, the numerical attribute "age" here is discretized before encoding, and any attribute $A_i$ without predicates is assigned the default value $\Delta_i$.

\vspace{1mm}
\noindent \textbf{Discretization.}
Discretization is essential for encoding numerical attributes like weight or salary, as their infinite possible values make direct encoding infeasible.
The discretization algorithm partitions the domain $\mathcal{D}$ of each numerical attribute into a predetermined number of bins,
and then, assigns values to their respective nearest bins.
This process aims to retain the key information of numerical data in the embedding space for preserving predictive capacity.

To this end, we employ a supervised discretization approach that takes into account the correlation between numerical attributes and the predictive target.
This method aims to maximize information value (IV), which measures the uncertainty reduction within each bin relative to the prediction target.
Higher IV values indicate a significant decrease in uncertainty, thereby preserving the predictive capacity.
Specifically, we introduce \textit{OptBinning}~\cite{navas2020optimal} implementation for discretization to optimize IV  while supporting constraints like the maximum bin count per attribute. The binning rule is learned on the training set and applied to the test set.
\label{subsec:SQLQueryModel}

\subsection{\Technique}
\label{sec:technique}

\begin{figure*}[t] 
\centering 
\includegraphics[width=0.8\textwidth]{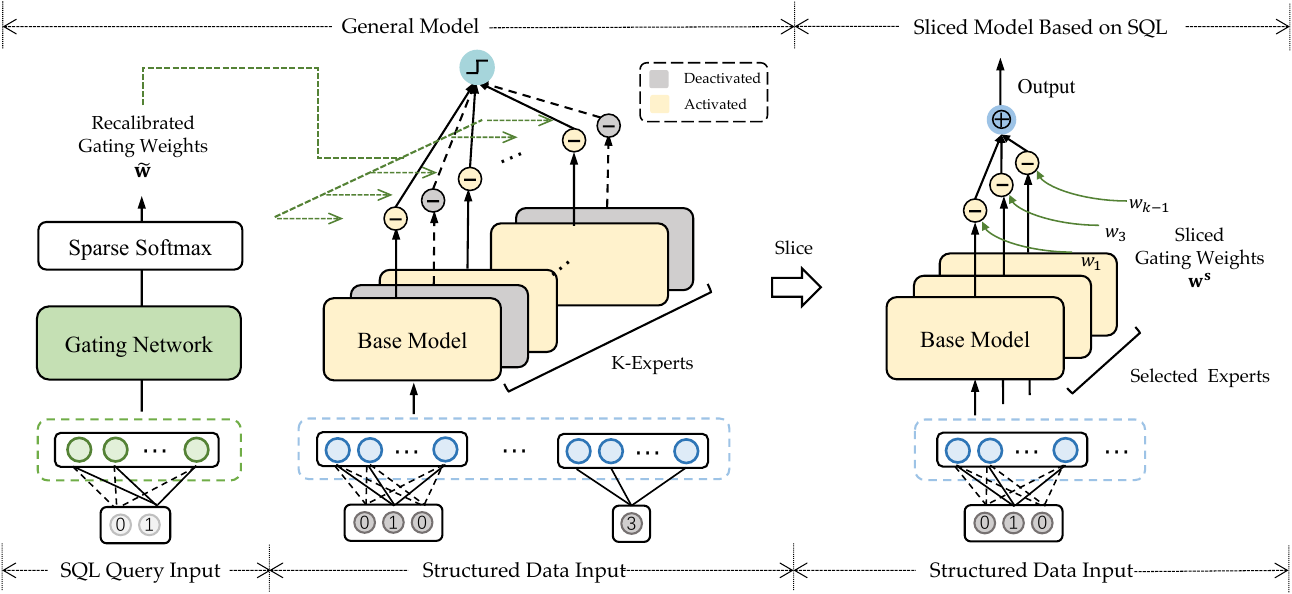} 
\caption{Overview of \technique.} 

\label{fig:network}
\end{figure*}

\bluerevise{
The SQL feature vector $\textbf{q}$, derived from the SQL query encoder, captures key information that can be exploited to tailor the predictive model for target subdatasets.
Our objective is to construct a general model with sufficient modeling capacity and then customize this model based on $\textbf{q}$ to enhance efficiency and predictive performance in SQL-aware predictive modeling.
While some approaches \cite{von2019hyper, wang2010evolving, liu2017learning} modify the model's architecture or parameters to maintain its size, they can lead to unstable predictive performance.
Conversely, ensemble approaches \cite{ganaie2022ensemble} improve performance by stacking models, but at the cost of additional computation.
To balance effectiveness and efficiency, we introduce \Name, which scales up the modeling capacity via MoE by replicating the base models as expert models, and integrates a SQL-aware gating network based on $\textbf{q}$ of SQL queries to selectively activate specific experts.
The overview of \Name is illustrated in Figure~\ref{fig:network}, and the key components following the data flow are introduced in this subsection.}

\noindent
\textbf{Preprocessing Module}
There are two sets of input constructed for the SQL-aware prediction modeling given an input tuple.
The first set of input is constructed for the gating network and can be uniformly represented as a categorical feature vector $\mathbf{q}$.
$\mathbf{q} = [q_1, q_2, \ldots, q_M]$ comprises $M$ feature values from respective attribute fields, where numerical attributes need to be converted into categorical attributes via discretization, as described in the previous subsection.
The second set is the attribute values of the input tuple $\mathbf{x} = [x_1, x_2, \ldots, x_M]$, and each attribute value $x_i$ can be either categorical or numerical.

For both $\mathbf{q}$ and $\mathbf{x}$, each field of attribute value $v_i$ ($q_i$/$x_i$) needs to be transformed into a corresponding embedding vector $\mathbf{e}_i$ to participate in the subsequent predictive modeling.
Specifically, each categorical attribute is transformed via \textit{embedding lookup}, i.e., $  \mathbf{e}_i =  \mathbf{E}_i[q_i], \mathbf{e}_i \in \mathbb{R}^{n_e}$, where $n_e$ is the dimension of embedding, and $\mathbf{E}_i$ is the embedding matrix of this categorical attribute.
Note that different embedding vectors in $\mathbf{E}_i$ correspond to their respective attribute values.
As for each numerical attribute $x_j$ in $\mathbf{x}$, the corresponding embedding vector is obtained by linearly scaling up a learnable embedding vector $\mathbf{\hat{e}}_j$
, i.e., $\mathbf{e}_j = x_j \cdot \mathbf{\hat{e}}_j$.
In this way, we can obtain fixed-size inputs, namely embedding vectors $\mathbf{\hat{q}} = [\mathbf{q}_1, \mathbf{q}_2, \ldots, \mathbf{q}_M]$ and $\mathbf{\hat{x}} = [\mathbf{x}_1, \mathbf{x}_2, \ldots, \mathbf{x}_M]$.

\vspace{1mm}
\noindent
\textbf{General Model and SQL-aware Gating Network.}
\Name replicates the base model $K$ times to construct the general model, denoted as $ \mathbf{\mathcal{F}} = [\mathcal{F}_1, \mathcal{F}_2, \ldots, \mathcal{F}_K]$
, where each base model is referred to as an ``expert model '' in MoE.
These expert models share the same architecture but learn distinct model parameters during training, which take the same input $\mathbf{\hat{x}}$ and produce different outputs that are later aggregated to form the final prediction.
The output of the $i$-th expert for a given input $\mathbf{x}$ is denoted as $\mathcal{F}_i (\mathbf{\hat{x}})$.

The SQL-aware gating network $\mathcal{G}$ takes the SQL query embedding vectors $\mathbf{\hat{q}}$ as input to produce a $K$-dimensional vector, termed the gating weight $\mathbf{w}$, with $\mathbf{w}\in \mathbb{R}^K $.
Specifically, a two-layer multilayer perceptron (MLP) is employed as the gating network following the practice~\cite{riquelme2021scaling,fedus2022switch,gpt4}.
We concatenate all embeddings in $\mathbf{\hat{q}}$ as the input of the gating network $\widetilde{\textbf{q}} = \mathbf{q}_1 \oplus  \mathbf{q}_2  \ldots \oplus \mathbf{q}_M$, where $\widetilde{\textbf{q}} \in \mathbb{R}^{M {\cdot}{n_e}} $,
then feed $\widetilde{\mathbf{q}}$ to $\mathcal{G}$, and obtain the gating weight $\mathbf{w}$ by:

\begin{equation}
\begin{aligned}
\mathbf{z} &= \phi(\mathbf{W_1}\widetilde{\mathbf{q}} + \mathbf{b_1}) \\
\mathbf{w} &=\mathcal{G}(\mathbf{s}) =  \mathbf{W_2}\mathbf{z} + \mathbf{b_2}
\end{aligned}
\end{equation}

\noindent
where $\mathbf{W_1} \in \mathbb{R}^{n_z \times Mn_e}, \mathbf{W_2} \in \mathbb{R}^{K \times n_z}$ and $\mathbf{b_1} \in \mathbb{R}^{n_z}, \mathbf{b_2} \in \mathbb{R}^{K}$  are the weights and biases respectively, $n_z$ is the hidden layer size, and $\phi$ represents the ReLU activation function.

Given the gating weight $\mathbf{w}$, the $\alpha$-entmax function~\cite{entmax, correia2019adaptively} is further applied to recalibrate $\mathbf{w}$ to a probability distribution.
As introduced in Section~\ref{sec:preliminaries}, the hyper-parameter $\alpha$ in $\alpha$-entmax controls the level of sparsity, and a larger value of $\alpha$ sets more gating weights to zero and thus deactivates more experts for higher efficiency.
The output of $\alpha$-entmax $\widetilde{\mathbf{w}}$ is:

\begin{equation}
    \widetilde{\mathbf{w}} = \alpha\text{-entmax}(\mathbf{w}),\  \widetilde{\mathbf{w}} \in \mathbb{R}^K 
\end{equation}

\noindent
which is utilized as weights to aggregate expert outputs.
The final output of the general model is a weighted average of expert outputs:

\begin{equation}
\label{equ:final_predict}
    \mathbf{\hat{y}} = \sum_{i=1}^K \widetilde{w}_i\cdot\mathcal{F}_i(\mathbf{x})
\end{equation}

\noindent
where $\mathbf{\hat{y}}$ is the prediction given the input $\mathbf{x}$ and the corresponding query $\mathbf{q}$ in the SQL-aware predictive modeling.

\vspace{1mm}
\noindent
\textbf{Dynamic Model Slicing via Gating Network.}
Given a SQL query, all retrieved data tuples share the same recalibrated gating weight $\widetilde{\mathbf{w}}$.
Notably, $\widetilde{w}_i = 0$ in Equation~\ref{equ:final_predict} indicates that the $i$-th expert is not required for the current current predictive task, and thus, only a selected subset of experts $\mathcal{F}_i$ are activated, ensuring a more efficient inference process.

Denoting the set of indices of activated experts as $\{ I_1, I_2,\cdots, I_{n_o} \}$, where $n_o$ is the current number of activated experts and $\widetilde{w}_{I_j} \neq 0, \forall j \in \{1, 2, \dots, n_o\}$, and given the corresponding SQL feature vector $\mathbf{q}$, we index the activated experts to form a sliced model, i.e., $\mathbf{\mathcal{F}}_{\mathbf{q}} = [\mathcal{F}_{I_1},\mathcal{F}_{I_2},\cdots,\mathcal{F}_{I_{n_o}}]$.
Thus, the final output is as follows:
\begin{equation}
    \mathbf{\hat{y}} = \sum_{j = 1}^{n_o} \widetilde{w}_{[I_j]} \cdot \mathcal{F}_{I_j}(\mathbf{x})
\end{equation}


\noindent
where the number of activated experts $n_o$ directly affects the effectiveness and efficiency of the sliced model.
A large $n_o$ indicates larger model capacity while incurring higher computational overhead, and vice versa.
In \Name, $n_o$ is determined by the gating network based on the SQL feature vector $\textbf{q}$ and the hyper-parameter $\alpha$ of the sparsemax function.
Notably, instead of predefining a fixed value, $\alpha$ in $\alpha$-entmax is learnable and optimized based on the input tuples and corresponding queries during training.
Subsequently, during inference, \Name can dynamically adapt $n_o$ based on the current SQL query, trading off between the effectiveness and efficiency of the predictive modeling.

\subsection{Optimization}
\label{sec:optimization}

Our \Name technique can be applied to different predictive tasks by configuring a proper objective function, such as mean squared error (MSE) for regression or cross-entropy for classification.
For example, in binary classification, the objective function employed is binary cross-entropy:

\begin{equation}
   {\rm LogLoss}({\hat{y}, y} ) = -\frac{1}{N} \sum_i^N \{ y_i  {\rm log} \sigma(\hat{y_i}) + 
    (1 - y_i) {\rm log} (1 - \sigma(\hat{y_i})) \}
\end{equation}

\noindent
where $\hat{y}$ is the prediction label, $y$ is the ground truth label, $N$ is the number of tuples, and $\sigma(\cdot)$ is the sigmoid function.

To make the optimization more robust and effective, we introduce two regularization terms into the main loss function.
The first term is the balance loss, $\mathcal{L}_{baln}$, which is to address \textit{imbalanced expert utilization}, a common issue in training MoE-based models.
This imbalance arises when the gating network $\mathcal{G}$ disproportionately favor a few experts, which skews the training dynamics.
As a consequence, preferred experts become overutilized while others remain underutilized, which compromises the MoE capacity and degrades overall model performance.


Let $\mathbf{X}$ denote a mini-batch of training instances with $n_b$ tuples, and
$\widetilde{\mathbf{W}} = [\widetilde{\mathbf{w}}_1, \widetilde{\mathbf{w}}_2, \cdots, \widetilde{\mathbf{w}}_{n_b}]$ represent the recalibrated gating weights of $\mathbf{X}$. 
Here, $\widetilde{w}_{ij}$ is the $j$-th weight of $\widetilde{\mathbf{w}}_{i}$.
$\mathcal{L}_{baln}$ is defined as follows:

\begin{equation}
\begin{aligned}
    \mathcal{L}_{baln} = {\rm cv} (\mathbf{\Phi}) &= 
    \sum_{j = 1}^{K}\frac{\phi_j - \mathbb{E}(\mathbf{\Phi})}{{\mathbb{E}(\mathbf{\Phi})}^2}\\
    \mathbf{\Phi} = [\phi_1, \phi_2, \cdots, &\phi_K], \space \phi_j =  \sum_{i=1}^{n_b} \widetilde{w}_{ij}
\end{aligned}
\end{equation}

\noindent
where $\mathbb{E}(\mathbf{\Phi})= \frac{1}{K} \sum_{j=1}^{K} \phi_j$.
This balance loss term promotes a uniform distribution of weights across experts in the mini-batch, encouraging equal contribution from each expert.
However, this term empirically tends to activate a large number of experts due to its drive for balance, which is contrary to the intent of sparse softmax to maintain sparsity.
To counterbalance this and enhance computational efficiency, we further introduce an additional term, the sparsity loss term, $\mathcal{L}_{sprs}$:

\begin{equation}
    \mathcal{L}_{sprs} = -\frac{1}{n_b} \sum_{i=1}^{n_b}{(\widetilde{\mathbf{w}}_i)}^2
\end{equation}

\noindent
which encourages the gating network to allocate higher weights on a select few experts while assigning minimal or zero weights to others.
Both loss terms are weighted by their respective \term{\textit{regularization coefficient}}, \(\lambda_1\) and \(\lambda_2\), and incorporated into the overall objective function:

\begin{equation}
  {\rm Loss}(\hat{\mathbf{y}}, {\mathbf{y}}) =
  {\rm LogLoss}(\hat{\mathbf{y}}, {\mathbf{y}}) + \lambda_1 L_{baln} + \lambda_2 L_{sprs}.
\end{equation}
\noindent
With this objective function, \Name can then be optimized effectively using 
gradient-based optimizers, e.g., SGD 
~\cite{amari1993backpropagation} 
or Adam~\cite{kingma2014adam}.

\begin{figure}
\centering
\includegraphics[width=0.49\textwidth]{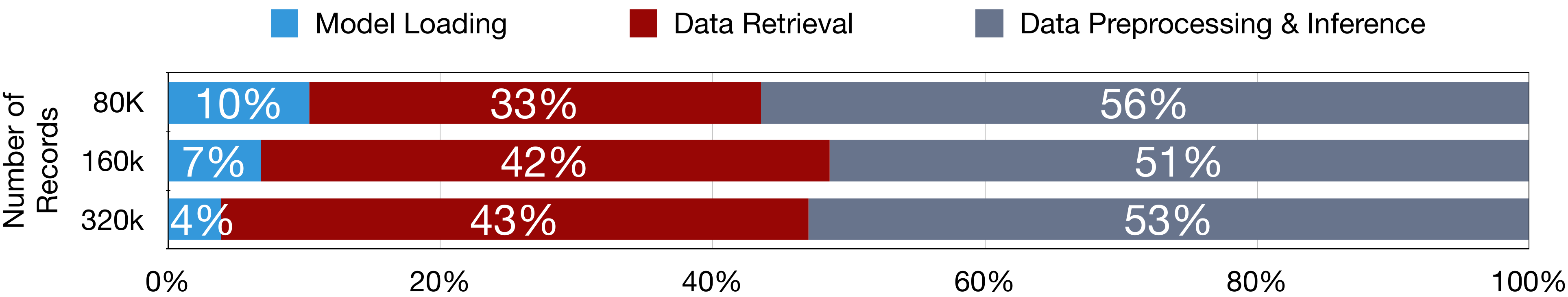}
\caption{Breakdown of inference response time for the Inference-Decouple Strategy.}
\label{fig:profile}
\end{figure}

\section{In-Database Model Inference}
\label{sec:indb_inference}
\red{
This section presents the implementation details of the in-database inference extension, which integrates the \Name technique into NeurDB to deliver more accurate predictions.
The typical inference pipeline consists of four stages:
(1) model loading, where the model is loaded into memory; 
(2) data retrieval, where subdatasets specified by SQL queries are fetched;
(3) data preprocessing, which transforms the raw data from the RDBMS into the tensor format for model inputs;
and (4) inference execution, where the model performs forward computation to generate predictions.
A straightforward approach to support \Name is to decouple the database and inference systems, referred to as the \textbf{Inference-Decouple Strategy} (IDS).}
In this approach, analysts retrieve subdatasets from the database, preprocess them, and run inference in a dedicated external inference system.
However, it has several drawbacks. First, exporting data from the database introduces security risks and may violate compliance regulations.
Second, managing two separate systems complicates the analytics workflow, increases operational complexity, and imposes additional learning requirements.
Third, transferring large volumes of data from the database to an external system incurs significant overhead and latency.
\red{Figure~\ref{fig:profile} illustrates the time usage breakdown of IDS
when running inference tasks.
Notably, data retrieval accounts for around 40\% of the total inference time primarily due to the overhead from database connections, data I/O, and network communication.
To address these issues, we utilize the \textbf{In-database Inference Strategy} (IIS), in which the database performs inference and avoids transferring data. 
This approach not only enhances security and simplifies workflow, but also accelerates the inference process. 
}

\begin{figure}
\centering
\includegraphics[width=0.46\textwidth]{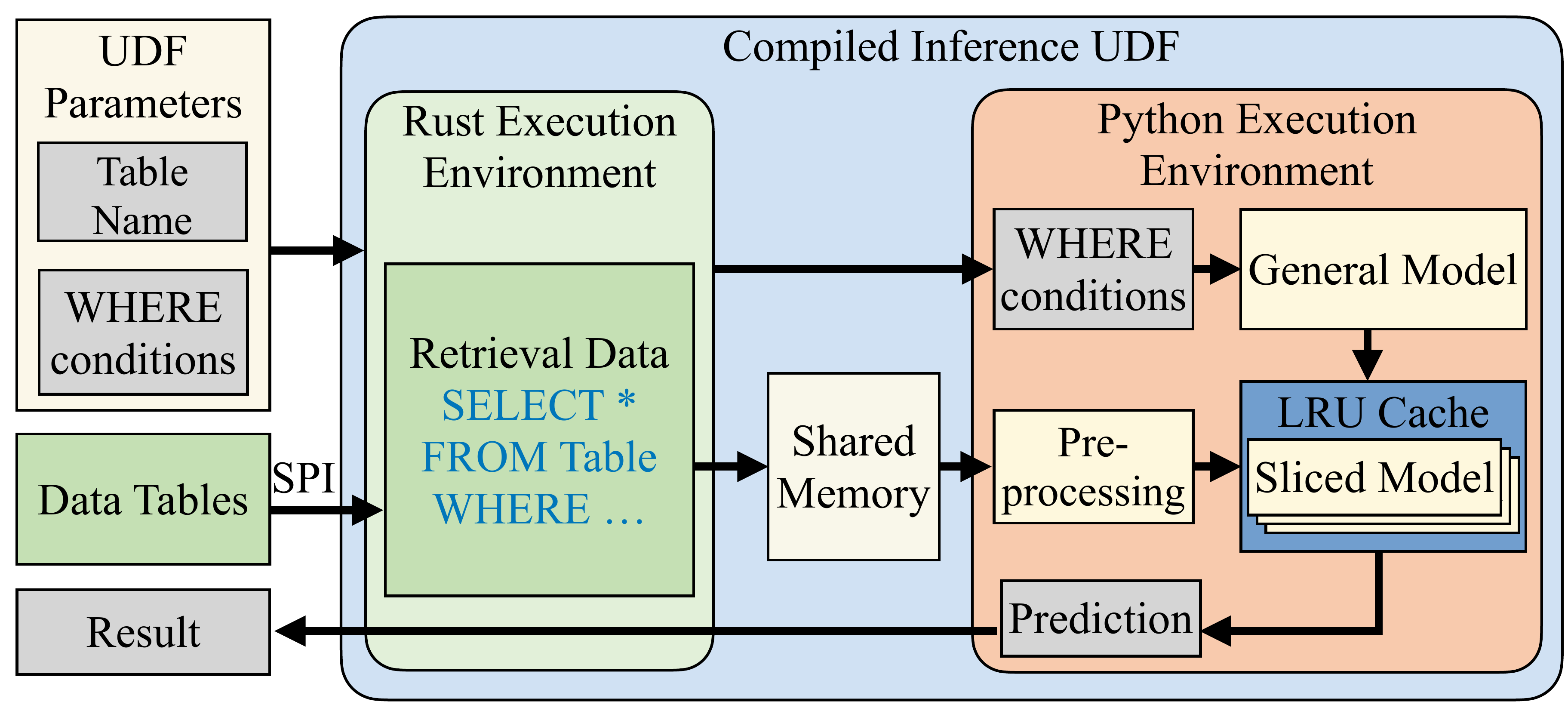}
\caption{Design and execution of inference UDF.}
\label{fig:udf}
\end{figure}

\begin{figure*}
\centering
\includegraphics[width=0.95\textwidth]{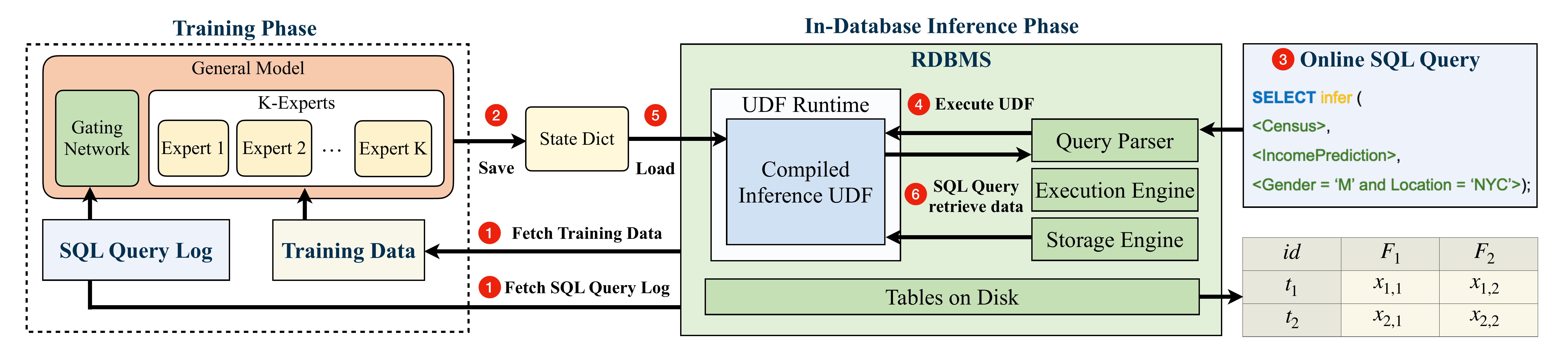}
\caption{\red{Workflow of in-database inference extension.}}
\label{fig:system}
\end{figure*}

\subsection{Inference UDF Design}
\label{subsec:indb_opt}
\red{
For the implementation, we design UDFs to encapsulate all stages of the inference process and integrate them into the database as an extension. 
As illustrated in Figure~\ref{fig:udf}., users begin by initiating predictive queries, specifying parameters `TableName' and `WHERE' conditions.
The UDF then retrieves relevant subdatasets by applying the `WHERE' condition on specified table, dynamically loads the trained model, and performs the inference.
Upon completion, the results are returned to the users.
However, directly embedding the decoupled inference process into UDFs is suboptimal due to Python's performance limitation. Therefore, we introduce three key optimizations to improve UDF execution efficiency.
}

\noindent
\highlight{Efficient Execution Allocation.} 
\red{
To optimize UDF performance, we adopt a multi-language strategy, combining the strengths of both Rust and Python. Rust is used for efficient data retrieval, leveraging the PostgreSQL extension development library PGRX~\cite{pgrx}. This allows us to utilize PostgreSQL’s built-in APIs, the Server Programming Interface (SPI) for high-performance data access. Meanwhile, Python is employed for data preprocessing and inference, benefiting from its compatibility with machine learning libraries like PyTorch and Scikit-learn. This combination ensures that we balance efficiency in data retrieval with the flexibility required for advanced data processing and machine learning tasks.
However, two main challenges remain for efficient inference:
(1) data copying overhead, caused by the need for extensive data transfers between different execution environments.
(2) state initialization overhead, resulting from the frequent loading of predictive models for each inference task.
To address these issues, we introduce memory sharing and state caching in our UDF designed to improve efficiency.}

\noindent
\highlight{Memory Sharing.} 
\red{
Data transfer between different execution environments requires two read-write operations: fetching data from RDBMS into Rust, then moving it to Python. To avoid this inefficiency, we introduce shared memory. Data is filtered in Rust using SPI and written directly to the shared memory, which is allocated before UDF invocation and accessible in both execution environment. The shared memory allows data to be used in Python without additional copying, eliminating unnecessary overhead.
}

\noindent
\highlight{State Caching.} 
\red{
As shown in Figure~\ref{fig:profile}, model loading takes up about 10\% of the total inference time per request. With many requests, the overhead of repeatedly loading the model grows. To reduce it, we implement session-level caching for the general model and maintain a state cache for the frequently used sliced models. 
Specifically, when processing a SQL query with a new filter condition, the UDF first checks the state cache for an existing sliced model. If none is found, a new sliced model is derived from the cached general model. We employ a Least Recently Used (LRU) cahcing strategy to manage the model cache, ensuring efficient memory use.
}

\subsection{System Workflow}
\label{sec:workflow}
\red{In this section, we describe the workflow of the in-database inference extension, which comprises both training and in-database inference phases as shown in Figure~\ref{fig:system}.}

\noindent
\highlight{Training Phase.} 
In the training phase, a general model is constructed according to the table schemas and the predictive task
SQL query logs from real-world scenarios are collected as the representative workload for training. These queries are utilized to retrieve the corresponding subdatasets, both of which are then preprocessed and fed into the general model for iterative training (Step 1 in Figure~\ref{fig:system}).
Upon completion of training, the general model is serialized and stored as a state dictionary (Step 2).
In response to the incoming SQL query, an associated UDF is invoked, and then the model is loaded into the database memory and sliced to handle various online inference requests.

\noindent
\highlight{In-Database Inference Phase.} 
The inference process is encapsulated into a UDF, which is integrated into database through extension installation.
The UDF, named \texttt{infer}, provides a SQL interface for inference queries, as shown in the statement below:
\begin{center}
\small
\begin{lstlisting}[style=SQL]
SELECT infer(<tableName>, <task>, <propositional formula>);
\end{lstlisting}
\end{center}
Here, \textit{<tableName>} indicates the source table for the subdataset.
\textit{<task>} specifies the prediction target (such as click-through rate or readmission rate) and directs to a particular deep learning model.
\textit{<propositional formula>} defines the filter conditions following the \texttt{WHERE} clause.
Upon receiving the inference query (Step 3 in Figure~\ref{fig:system}), the UDF is executed as illustrated in Figure~\ref{fig:udf} (Step 4).
The UDF performs four main tasks for an online inference request:
(1) The prediction target and model are specified by \textit{<tableName>} and \textit{<task>}. If the model is not cached, the UDF will locate and load the general model (Step 5).
(2) The UDF slices this model according to the \textit{<propositional formula>} under the guidance of the \Name technique.
(3) Predictive data is retrieved via the SPI and written into shared memory (Step 6).
(4) The sliced model is applied to the data for inference.
Finally, the UDF returns a view containing the original data alongside a new predictive column.

\section{Experiments}
\label{sec:experiment}
In this section, we evaluate the effectiveness of  \Name~ and efficiency of our \red{in-database inference extension}, using \bluerevise{five} real-world datasets.
We first introduce the datasets and experimental setup, and then design experiments and report findings to address four key research questions (RQs):
\begin{itemize}
    \item RQ1: Does the \Name~ technique improve the SQL-aware predictive modeling task compared to base models?
    \item RQ2: How effective is each component in \Name technique?
    \item RQ3: \red{Does our in-database inference extension improve the efficiency compared to traditional approaches?}
    \item RQ4: \red{How effective is our extension in complex scenarios?}
\end{itemize}

\subsection{Experimental Setup}
\label{subsec:expSetup}
\subsubsection{Datasets} 
\label{subsubsec:dataset} 
We conduct experiments on five real-world datasets from the domains of finance, sociology, and healthcare.
The statistics of the datasets are summarized in Table~\ref{tab:dataset}.
\\\noindent (1) \textbf{Payment} \cite{payment, payment_source} consists of credit card clients' profiles and their past bill payments. The task is to predict if a credit card payment will default next month.
\\\noindent(2) \textbf{Credit} \cite{credit, credit_source} is gathered by Home Credit Group, focusing on the unbanked population. The task is to predict the repayment abilities of this population for better loan experience. 
\\\noindent(3) \textbf{Census} \cite{census, census_source} contains data from the Current Population Survey conducted by the U.S. Census Bureau. The task is to determine whether a person's annual income exceeds 50K based on their profile information, including age, class education, etc.
\\\noindent(4)  \textbf{Diabetes} \cite{diabetes, diabetes_source} comprises ten years of clinical care at 130 US hospitals. Each tuple pertains to hospital records of patients diagnosed with diabetes, including details like medications and laboratory results. The task is to predict the patient's readmission.
\bluerevise{
\\\noindent(5)
\textbf{Avazu} \cite{avazu} collects data from a mobile platform to predict click-through rates on ads.
The dataset comprises millions of records and 22 attribute fields, covering mobile application and device information, with a total of 1,544,250 unique features.
}

\subsubsection{Workloads} 


\begin{table}[]
\centering
\small
\setcounter{table}{0}
\caption{\revision{Dataset statistics.}}
\label{tab:dataset}
\resizebox{.95\columnwidth}{!}{%
\begin{tabular}{lcccc}
\hline
Dataset & Tuples   & Positive Ratio & Attributes & Features \\ \hline
Payment   & 30,000  & 21.4\%     & 23     & 350     \\
Credit    & 244,280  & 7.8\%      & 69     & 550      \\
Census & 269,356  &  6.4\%     & 41     & 540      \\
Diabetes  & 101,766 & 46.8\%    & 48    & 850     \\ 
Avazu & 40,428,967 & 17.2\% &22 & 1,544,250\\ \hline  
\end{tabular}%
}
\end{table}

\label{subsubsec:workload}
In SQL-aware predictive modeling, traditional OLAP benchmarks like TPC-DS~\cite{nambiar2006making} and TPC-H focus on evaluating query performance using operations such as \texttt{JOIN} and \texttt{GROUPBY}, which lack prediction tasks.
In contrast, existing prediction datasets used for evaluating predictive modeling performance do not incorporate SQL queries.
To address this deficiency, we develop a method to sample synthetic \textit{inference queries} for the evaluation of SQL-aware predictive modeling.

Our method, detailed in Algorithm~\ref{alg:create_workload}, employs a random strategy to create a set of synthetic SQL queries. The generation procedure is as follows:
(1) randomly select a data tuple $\textbf{x}$ from the dataset $D$ (Step 3);
(2) sample a value $m$ as the number of predicates in the current query from the range $[1, \min(max\_{col}, M)]$, where $max\_{col}$ is the maximum number of predicates allowed in SQL queries, and $M$ is the number of attributes in $D$ (Step 4);
(3) sample $m$ attributes from $\textbf{x}$ and use their values to construct a propositional formula for the SQL query (Steps 5-6);
(4) add the generated SQL query to the workload (Step 7).
This procedure is repeated $N$ to create a comprehensive workload.
In experiments, we set $N$ to 50 and $max\_col$ to 3, and generate workloads on each dataset accordingly.

\subsubsection{Baseline Methods}
\label{sec:exp_baseline}
We select four kinds of base predictive models and enhance these models via the \Name technique. We evaluate \Name's effectiveness by comparing the performance of these base models with and without the integration of \Name. The base models are as follows
\\\noindent (1) \textbf{DNN} \cite{gardner1998artificial}:  contains fully-connected linear and activation layers, representing the most fundamental neural network. 
\\\noindent (2) \textbf{CIN} \cite{lian2018xdeepfm}: it models higher-order feature combinations through compressed interaction with input embeddings.
\\\noindent (3) \textbf{AFN} \cite{cheng2020adaptive}: it incorporates logarithm neurons in the network layer, aiding in capturing the feature interaction in arbitrary order.
\\\noindent (4) \textbf{ARMNet} \cite{cai2021arm}: it introduces multi-head attention to adaptively extract the combination of features, demonstrating state-of-the-art performance in structured data prediction tasks.

\red{
Additionally, to evaluate the efficiency of the in-database inference extension, we compare it with the traditional analytic approach IDS
described in Section~\ref{sec:indb_inference}, in which data is retrieved out of the database through network communication based on \textit{psycopg}, while no data is copied between execution environments.
} 

\begin{algorithm}[t]
\caption{\bluerevise{Synthetic Workload Generation}}
\label{alg:create_workload}
\begin{algorithmic}[1]
\Require dataset $D$, the number of SELECT queries $N$, the maximum filter condition size $max\_col$
\Ensure a synthetic workload $W$ containing $N$ SELECT queries
\State $W = \varnothing$
\For{$i \gets 1$ to $N$} 
    \State Randomly select a data tuple  $\mathbf{x} \in \mathbb{R}^M$ from $D$ 
    \State Randomly sample the number of selected columns $m \in [1, \min(max\_col, M)]$
    \State Randomly sample $m$ columns from data tuple $\mathbf{x}$ along with their corresponding values
    \State Form a SELECT query with a filter condition of size $m$ based on the selected columns and values
    \State Add the generated SELECT query to the  workload $W$ 
\EndFor

\State \textbf{return} synthetic workload $W$
\end{algorithmic}
\end{algorithm}

\begin{table*}[t]
\centering
\caption{Evaluation of performance improvements with LEADS.}
\label{tab:mainstudy}
\resizebox{\textwidth}{!}{%
\begin{tabular}{@{}lccccccccccccc@{}}
\toprule[1.5pt]
                           &                          & \multicolumn{3}{c}{DNN}                                   & \multicolumn{3}{c}{CIN}                                  & \multicolumn{3}{c}{AFN}                                                         & \multicolumn{3}{c}{ARMNet}          \\
\multirow{-2}{*}{Datasets} & \multirow{-2}{*}{Metric} & w/o    & w/              & \multicolumn{1}{c|}{Imprv.}    & w/o    & w/              & \multicolumn{1}{c|}{Imprv.}   & w/o    & w/                                     & \multicolumn{1}{c|}{Imprv.}   & w/o    & w/              & Imprv.   \\ \midrule
                           & Workload-AUC             & 0.7003 & \textbf{0.7089} & \multicolumn{1}{c|}{+1.23\%}   & 0.7164 & \textbf{0.7189} & \multicolumn{1}{c|}{+0.35\%}  & 0.7067 & \textbf{0.7143}                        & \multicolumn{1}{c|}{+1.08\%}  & 0.7141 & \textbf{0.7212} & +0.99\%  \\
\multirow{-2}{*}{Payment}  & Worst-AUC                & 0.4733 & \textbf{0.5467} & \multicolumn{1}{c|}{+15.51\%}  & 0.3836 & \textbf{0.4463} & \multicolumn{1}{c|}{+16.35\%} & 0.4467 & \textbf{0.6333}                        & \multicolumn{1}{c|}{+41.77\%} & 0.5267 & \textbf{0.6067} & +15.19\% \\ \midrule
                           & Workload-AUC             & 0.7145 & \textbf{0.7427} & \multicolumn{1}{c|}{+3.95\%}   & 0.7234 & \textbf{0.7408} & \multicolumn{1}{c|}{+2.41\%}  & 0.7171 & \textbf{0.7218}                        & \multicolumn{1}{c|}{+0.66\%}  & 0.7231 & \textbf{0.7347} & +1.60\%  \\
\multirow{-2}{*}{Credit}   & Worst-AUC                & 0.3852 & \textbf{0.6000} & \multicolumn{1}{c|}{+55.76\%} & 0.3333 & \textbf{0.4074} & \multicolumn{1}{c|}{+22.23\%} & 0.3852 & \textbf{0.4074}                        & \multicolumn{1}{c|}{+5.76\%}  & 0.4444 & \textbf{0.6264} & +40.95\% \\ \midrule
                           & Workload-AUC             & 0.9157 & \textbf{0.9200} & \multicolumn{1}{c|}{+0.47\%}   & 0.9187 & \textbf{0.9224} & \multicolumn{1}{c|}{+0.40\%}  & 0.9151 & \textbf{0.9216}                        & \multicolumn{1}{c|}{+0.71\%}  & 0.9196 & \textbf{0.9237} & +0.45\%  \\
\multirow{-2}{*}{Census}   & Worst-AUC                & 0.7692 & \textbf{0.8041} & \multicolumn{1}{c|}{+4.54\%}   & 0.7692 & \textbf{0.7845} & \multicolumn{1}{c|}{+1.99\%}  & 0.7577 & {\color[HTML]{000000} \textbf{0.7892}} & \multicolumn{1}{c|}{+4.16\%}  & 0.7692 & \textbf{0.7962} & +3.51\%  \\ \midrule
                           & Workload-AUC             & 0.8308 & \textbf{0.8375} & \multicolumn{1}{c|}{+0.81\%}   & 0.8322 & \textbf{0.8419} & \multicolumn{1}{c|}{+1.17\%}  & 0.8329 & \textbf{0.8390}                        & \multicolumn{1}{c|}{+0.73\%}  & 0.8342 & \textbf{0.8402} & +0.72\%  \\
\multirow{-2}{*}{Diabetes} & Worst-AUC                & 0.5495 & \textbf{0.6374} & \multicolumn{1}{c|}{+16.00\%}     & 0.6264 & \textbf{0.7033} & \multicolumn{1}{c|}{+12.28\%} & 0.6484 & \textbf{0.6813}                        & \multicolumn{1}{c|}{+5.07\%}  & 0.6044 & \textbf{0.6593} & +9.08\%  \\  \midrule
                        & Workload-AUC             & 0.7355 & \textbf{0.7424} & \multicolumn{1}{c|}{+0.94\%}   & 0.7324 & \textbf{0.7443} & \multicolumn{1}{c|}{+1.62\%}  & 0.7364 & \textbf{0.7424}                        & \multicolumn{1}{c|}{+0.81\%}  & 0.7387 & \textbf{0.7440} & +0.71\%  \\
\multirow{-2}{*}{Avazu} & Worst-AUC                & 
0.4231
& \textbf{0.5562} & \multicolumn{1}{c|}{+31.46\%}     & 
0.4531
& \textbf{0.5625} & \multicolumn{1}{c|}{+24.14\%} & 
0.5031
& \textbf{0.5594}                        & \multicolumn{1}{c|}{+11.19\%}  & 
0.4615
& \textbf{0.5625} & +21.88\%  \\ \bottomrule[1.5pt]
\end{tabular}%
}
\end{table*}

\subsubsection{Evaluation Metric} \label{subsubsec:metric}
A workload consists of a set of inference queries, each representing a prediction request. We evaluate the effectiveness of \Name on a single inference query, denoted by $q$, using the AUC (Area Under the ROC Curve) metric, where higher values indicates better performance.
To assess \Name's overall performance across the entire workload, we use two metrics. The first is the average AUC for all queries, denoted as \textit{Workload-AUC}:
\begin{align}
\label{eq:workload-auc}
 Workload\text{-}AUC(W) = \frac{1}{N} \sum_{i = 0}^N AUC(q_i)
\end{align}
where $N$ is the number of SQL queries, $q_i$ is the $i$-th query in the workload $W$.
The second is the lowest AUC value among all queries, termed \textit{Worst-AUC} and calculated as:
\begin{align}
\label{eq:workload-auc}
   Worst\text{-}AUC(W) = Min\{AUC(q_1), AUC(q_2), \dots ,AUC(q_N)\}
\end{align}
In fields like finance and healthcare where mistakes can lead to significant losses, it's important to focus on the worst-case performance. The Worst-AUC reveals whether the technique performs reliably and helps avoid poor decisions.
To assess model-level efficiency, we utilize the floating-point operations per second (\textit{FLOPs}) to measure the computations during the inference phase.
For system-level evaluation, we measure the performance using \textit{response time}, which tracks the CPU time elapsed from when a user initiates a query to when the prediction results are received.

\subsubsection{Hyper-parameter Settings} \label{subsubsec:hyperparameter}
For fair comparisons, we set the feature embedding size to 10 and the hidden layer size to 32 for all base models. Since \Name allows multiple experts, we reduce the hidden layer size of each expert to 16 for efficiency. For ARMNet, we set the self-attention module to 4 heads and a hidden size of 16. The initial $\alpha$ in sparsemax is 2.5. The number of experts in \Name is selected between 2 and 256 and fixed at 16. The balance regularization factor $\lambda_1$ and sparsity factor $\lambda_2$ are chosen between 1e-3 and 5e-2, both fixed at 1e-3. We conduct a sensitivity analysis on these hyperparameters and report the best results.

\subsubsection{Training Details}
Since the training dataset lacks specific SQL queries to update the gating network parameters, we simulate a SQL query for each input following Steps 3-6 in Algorithm~\ref{alg:create_workload}
. We use the Adam optimizer~\cite{kingma2014adam} with a learning rate range of 1e-3 to 0.1 and a batch size of 1024 for all base models and datasets.
Experiments are conducted on a server with a Xeon Silver 4114 CPU @ 2.2GHz (10 cores), 256GB of memory, and a GeForce RTX 3090 Ti. All models are implemented using PyTorch 1.6.0 with CUDA 10.2.


\begin{figure}[t]
\small
\centering
\includegraphics[width=0.9\columnwidth]{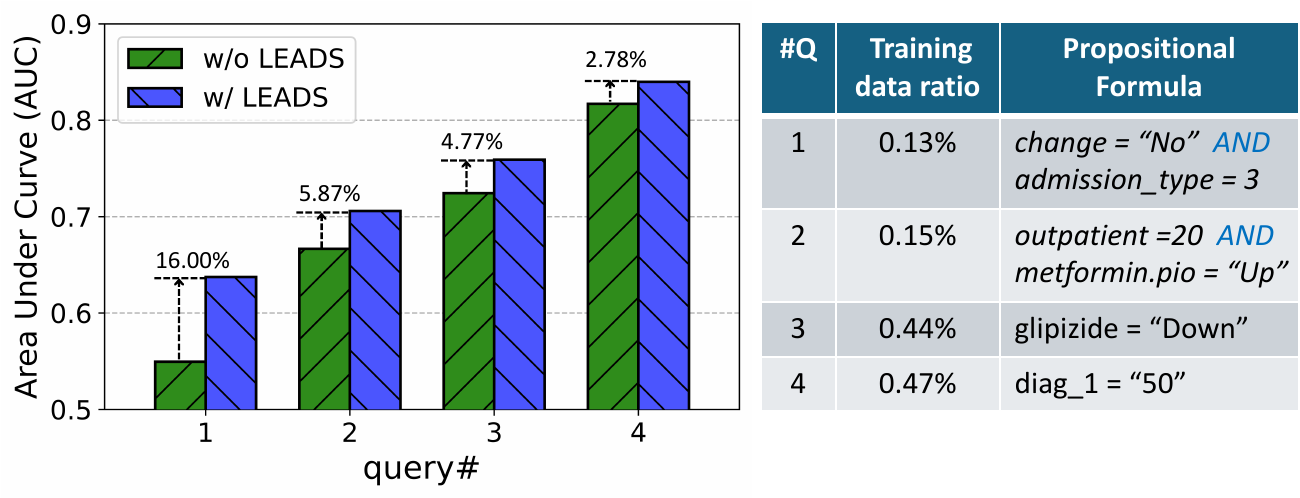} 
\captionsetup{justification=centering}
\caption{
\bluerevise{Top-4 SQL queries in terms of AUC improvement.}}
\label{fig:worst_sql_compare}
\end{figure}

\subsection{SQL-aware Predictions} 
\label{subsec:RQ1}
To answer the question RQ1, we investigate the performance of four base models with \Name. and the results are summarized in Table~\ref{tab:mainstudy}.
The main observation is that the prediction performance w.r.t. \textit{Workload-AUC} and \textit{Worst-AUC} consistently improve when utilizing \Name on base models 
Notably, the most significant improvement is observed in the Worst-AUC metric.
For instance, when using DNN as the base model, \Name achieves improvements of 55.76\% and 16.00\% on the Credit and Diabetes datasets, respectively.
The reason for the base model's low performance 
could be significant variability or nuances in the instances of the retrieved subset that are not well-represented in the training data. Consequently, the trained base model fails to provide accurate predictions.

To further analyze the Worst-AUC improvement, we perform a breakdown analysis on the Diabetes dataset with the DNN base model. Figure~\ref{fig:worst_sql_compare} presents the top-4 SQL queries with the highest AUC improvements from \Name, along with their query profiles.
We notice that these queries have lower AUC values compared to the overall Workload-AUC (see Table~\ref{tab:mainstudy}) and involve small subsets of the training data.
For instance, query\#1 uses only 0.13\% of the training data, about 130 out of 101,766 tuples. With such limited data, the base model struggles to generalize, resulting in poor predictions. \Name addresses this by leveraging the propositional formulas in SQL queries to help the general model identify patterns in these small subdatasets, enhancing performance on queries with few training samples. In Diabetes dataset, many queries involve the drug status. For example, query\#3 indicates a reduced dosage of \textit{glipizide}, which significantly impacts readmission rates. This demonstrates \Name's value in improving healthcare predictions.

\begin{figure}[t]
\centering
\begin{subfigure}[b]{0.49\textwidth}
    \hspace{1mm}
    \centering
    \includegraphics[width=0.93\columnwidth]{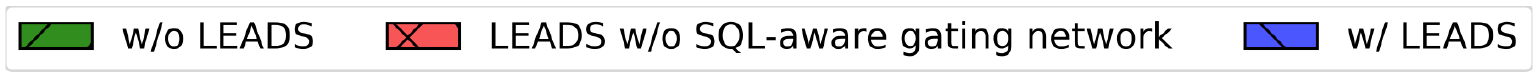}
\end{subfigure}

\begin{subfigure}[b]{0.235\textwidth}
    \centering
    \captionsetup{justification=centering,margin={0.5cm,0cm}}
    \includegraphics[width=0.95\columnwidth]{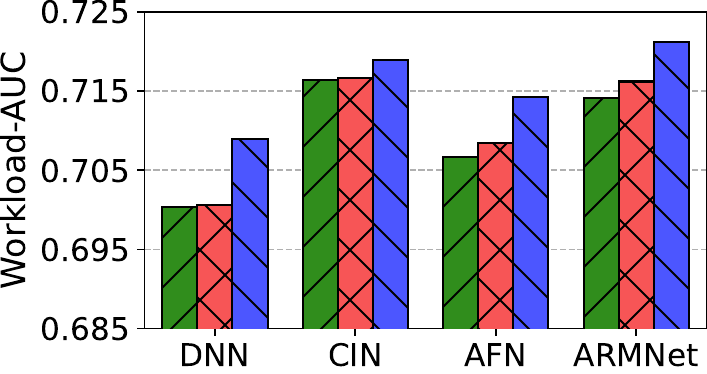}
    \caption{Payment.}
\end{subfigure}
~
\begin{subfigure}[b]{0.235\textwidth}
    \centering
    \captionsetup{justification=centering,margin={0.5cm,0cm}}
    \includegraphics[width=0.95\columnwidth]    {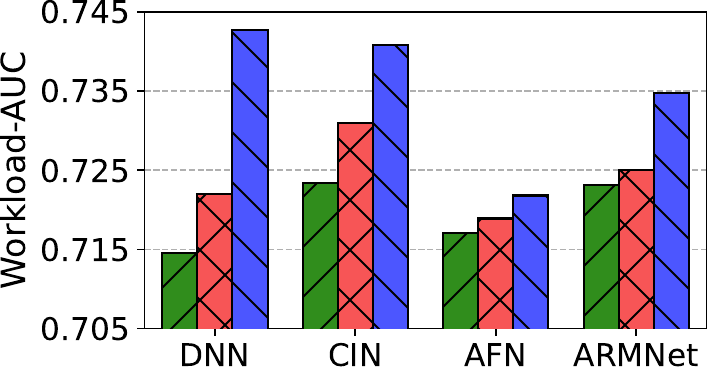}
    \caption{Credit.}
\end{subfigure}
\caption{Effects of SQL-aware gating network on accuracy.} 
\label{fig:sqlgate_ablation}
\end{figure}
\begin{table}[]
\caption{\bluerevise{Effects of the number of predicates on AUC.}}
\label{tab:query_complexity_revision}
\resizebox{\columnwidth}{!}{%
\begin{tabular}{@{}cccccccc@{}}
\toprule
Dataset                 & \#predicates & 1      & 2      & 3      & 4      & 5      & 6      \\ \midrule
\multirow{3}{*}{Credit} & \#tuple      & 24375  & 9902   & 9900   & 9896   & 9856   & 5208   \\ \cmidrule(l){2-8} 
                        & w/o LEADS    & 0.7432 & 0.7495 & 0.7495 & 0.7494 & 0.7483 & 0.7494 \\
 & w/ LEADS & \textbf{0.7522} & \textbf{0.7589} & \textbf{0.7591} & \textbf{0.7581} & \textbf{0.7577} & \textbf{0.7604} \\ \bottomrule
\end{tabular}%
}
\end{table}



\begin{figure*}[t]
\centering
\begin{subfigure}[b]{0.49\textwidth}
    \hspace{1mm}
    \centering
    \includegraphics[width=0.75\columnwidth]{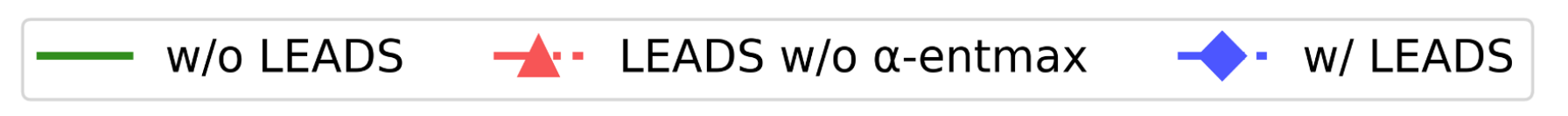}
\end{subfigure}

\begin{subfigure}[b]{0.24\textwidth}
    \centering
    \captionsetup{justification=centering,margin={0cm,0.7cm}}
    \includegraphics[width=0.99\columnwidth]{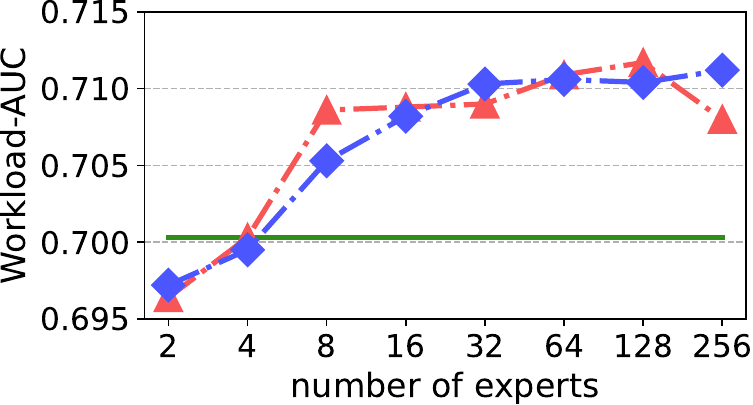}
        \caption{Payment-AUC.}
    \label{fig:sql}
\end{subfigure}
~
\begin{subfigure}[b]{0.24\textwidth}
    \centering
    \captionsetup{justification=centering,,margin={0cm,0.7cm}}
    \includegraphics[width=0.99\columnwidth]{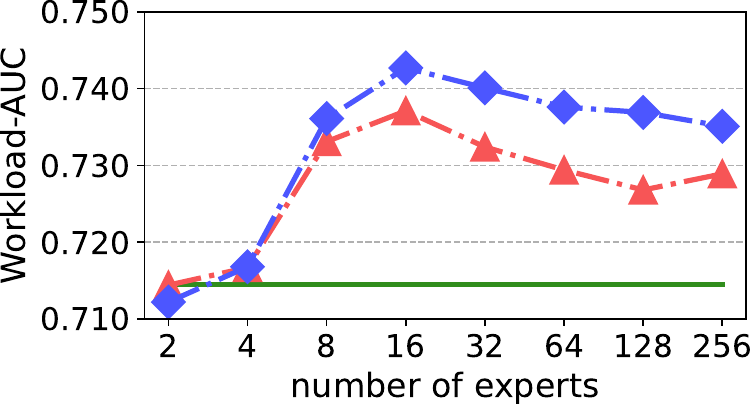}
    \caption{Credit-AUC.}
    \label{fig:db_scalability}
\end{subfigure}
~
\begin{subfigure}[b]{0.24\textwidth}
    \centering
    \includegraphics[width=0.97\columnwidth]{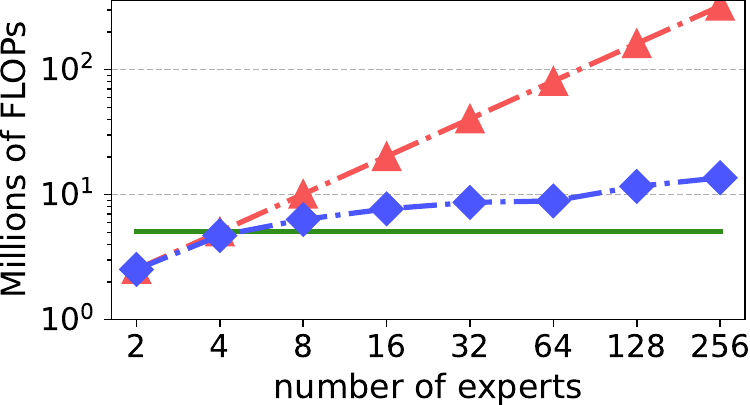}
        \caption{Payment-FLOPs.}
    \label{fig:sql}
\end{subfigure}
~
\begin{subfigure}[b]{0.24\textwidth}
    \centering
    \captionsetup{justification=centering,margin={0cm,0.7cm}}
    \includegraphics[width=0.97\columnwidth]{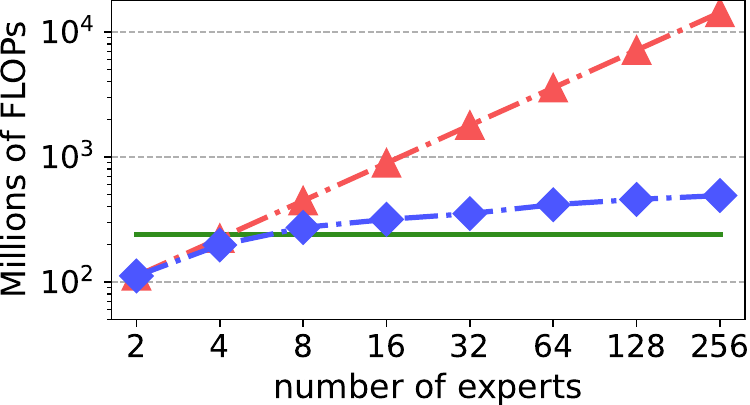}
    \caption{Credit-FLOPs.}
    \label{fig:db_scalability}
\end{subfigure}
\caption{Effects of $\alpha$-entmax and number of experts.}
\label{fig:exp_sparsemax_auc}
\end{figure*}

We also investigate the impact of predicate numbers on predictive results, as it represents query complexity. 
In our experiment, we incrementally added up to 6 predicates to evaluate the prediction. As shown in Table~\ref{tab:query_complexity_revision}, \Name outperforms baseline methods across different levels of query complexity. Additionally, there is an upward trend in predictive performance with the increase of the predicate number, indicating that complex primitive queries provide richer meta-information and enhance prediction accuracy.

\begin{figure}[t]
\centering
\captionsetup{justification=centering}
\begin{subfigure}[b]{0.45\textwidth}
    \hspace{1mm}
    \centering
    \includegraphics[width=0.8\columnwidth]{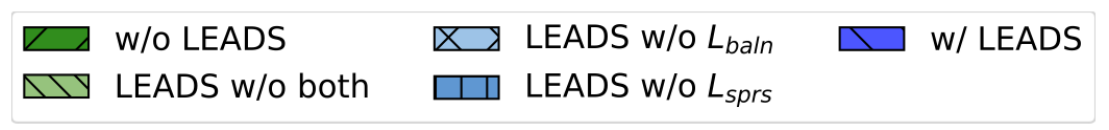}
\end{subfigure}
\vspace{1mm}

\begin{subfigure}[b]{0.235\textwidth}
    \centering
    \captionsetup{justification=centering,margin={0cm,0.5cm}}
    \includegraphics[width=0.96\columnwidth]{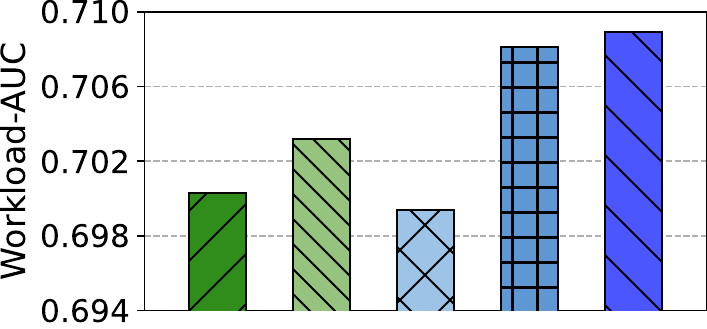}
    \caption{Payment.}
    \label{fig:sql}
\end{subfigure}~
\begin{subfigure}[b]{0.235\textwidth}
    \centering
    \captionsetup{justification=centering,margin={0cm,0.5cm}}
    \includegraphics[width=0.96\columnwidth]{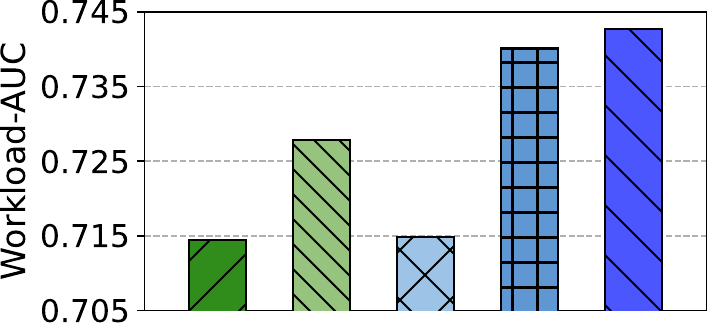}
    \caption{Credit.}
    \label{fig:db_scalability}
\end{subfigure}
\caption{Effects of the 
regularization terms on accuracy.
}
\label{fig:exp_reg_auc}
\end{figure}

\subsection{Ablation Study}
\label{subsec:RQ2}
In this part, we conduct the ablation study 
to answer the question RQ2, evaluating the effectiveness of each component in \Name.

\vspace{1mm}
\noindent \textbf{SQL-aware gating network}. In this evaluation, we remove the \textit{SQL-aware gating network} to demonstrate the importance of dynamic expert selection. A default model is created using a special SQL query embedding, where a set of padding values for each attribute, denoted as $\textbf{q}_d = [\Delta_1, \Delta_2, \cdots, \Delta_M ]$, indicating the absence of predicates in the SQL query.
The comparison results are shown in Figure~\ref{fig:sqlgate_ablation}. 
There are two main observations. First, the \textit{w/o \Name} method achieves the lowest Workload-AUC because it simply uses the base model to handle all SQL queries.
Second, the \textit{\Name w/o SQL-aware gating network} method results in a performance reduction compared to \Name. For example, on the Credit dataset, this reduction can reach up to 0.02 in terms of Workload-AUC. It is because without the SQL-aware gating network, \Name loses the ability for dynamic model customization based on SQL query vectors, leading to unsatisfactory results.

\vspace{1mm}
\noindent \textbf{$\alpha$-entmax}.
We evaluate the effect of the \textit{$\alpha$-entmax} function by comparing \Name to \textit{w/o \Name} and \textit{\Name w/o $\alpha$-entmax} (which uses the softmax function)
Using DNN as the base model, we vary the number of experts from 2 to 256 and measure the performance w.r.t. Workload-AUC and FLOPs on Payment and Credit datasets. 
As shown in Figures~\ref{fig:exp_sparsemax_auc}, increasing the number of experts from 2 to 256 leads to notable AUC improvements, as more experts allow the model to make more accurate predictions.
However, beyond 32 experts, \Name w/o $\alpha$-entmax sees a decline in AUC on the Credit dataset due to easily overfitting, as the model becomes too complex.
Figure~\ref{fig:exp_sparsemax_auc} also highlights the advantages of $\alpha$-entmax in terms of FLOPs saving.
Specifically, the FLOPs of \Name w/o $\alpha$-entmax grow linearly with more experts, while \Name with $\alpha$-entmax has a gentler increase, because $\alpha$-entmax assigns small values to zero, reducing the number of active experts compared to softmax. We remove these unused experts to conserve computational resources.

\begin{figure}[t]
\centering
\includegraphics[width=\columnwidth]{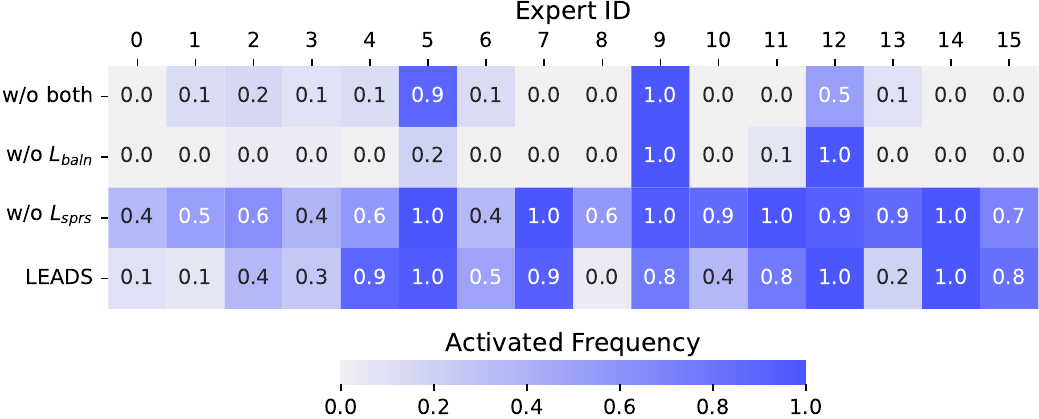} 
\caption{
Analysis of the activated frequency for each expert. The value denotes the frequency and 1.0 means the expert is activated in every SQL query. 
}
\label{fig:exp_reg_expert_frequency}
\end{figure}




\begin{figure*}[t]
\centering
\begin{subfigure}[b]{0.7\textwidth}
    \hspace{1mm}
    \centering    \includegraphics[width=0.95\columnwidth]{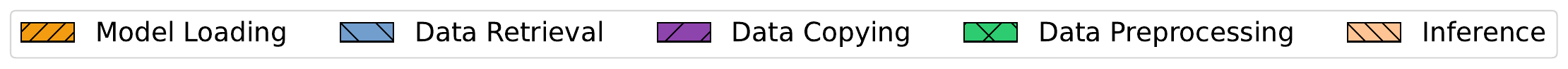}
\end{subfigure}

\begin{subfigure}[b]{0.23\textwidth}
    \centering
    \includegraphics[width=0.95\columnwidth]{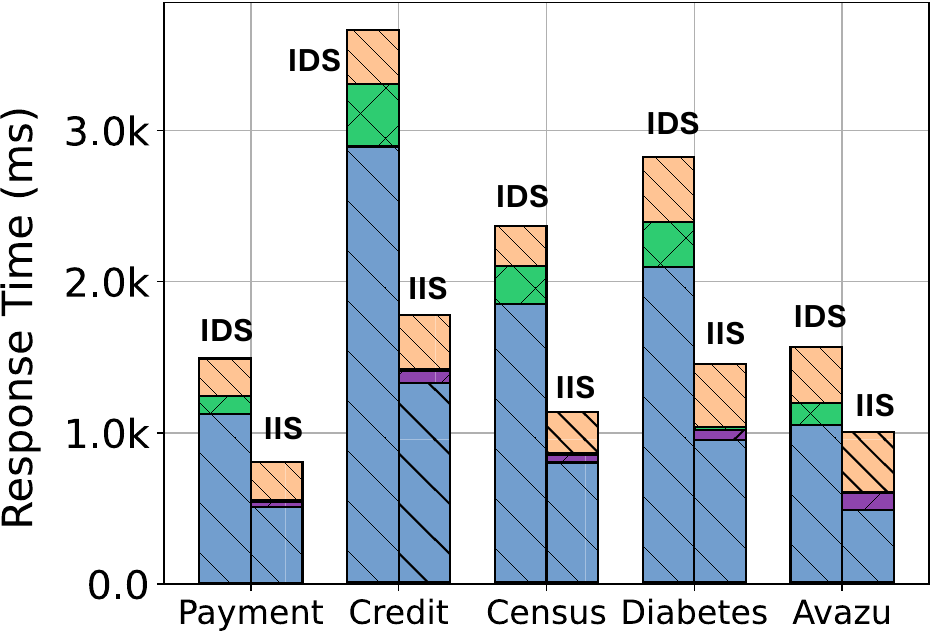}
            \caption{Response time for predicting 100k records on five datasets.}
    \label{fig:db_macro}
\end{subfigure}
\hspace{0.1mm}
\begin{subfigure}[b]{0.23\textwidth}
    \centering
    \includegraphics[width=0.95\columnwidth]{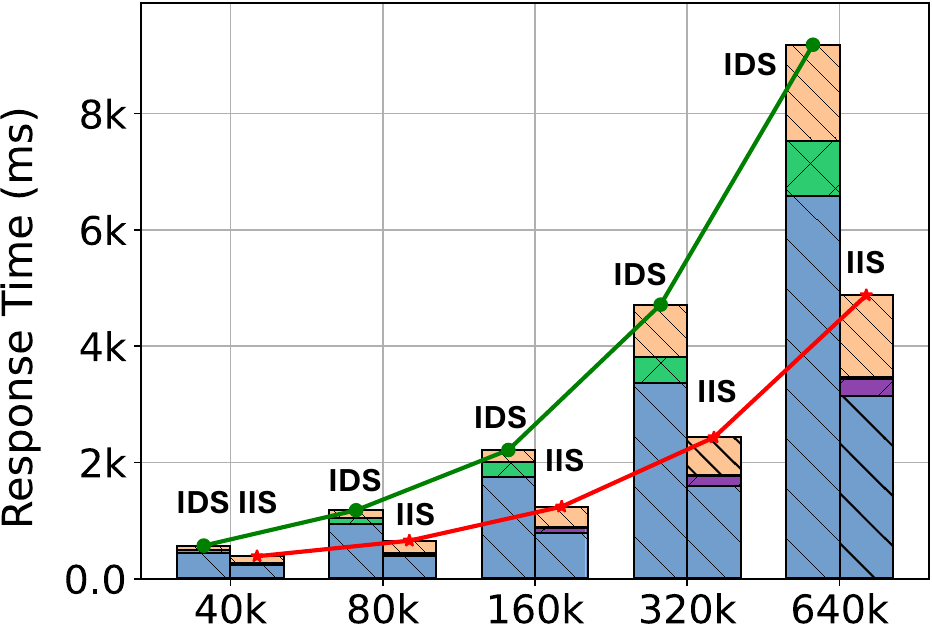}
    \caption{Response time w.r.t. \#predicting records on Payment dataset.}
    \label{fig:db_scalability}
\end{subfigure}
\hspace{0.1mm}
\begin{subfigure}[b]{0.23\textwidth}
    \centering
    \includegraphics[width=0.95\columnwidth]{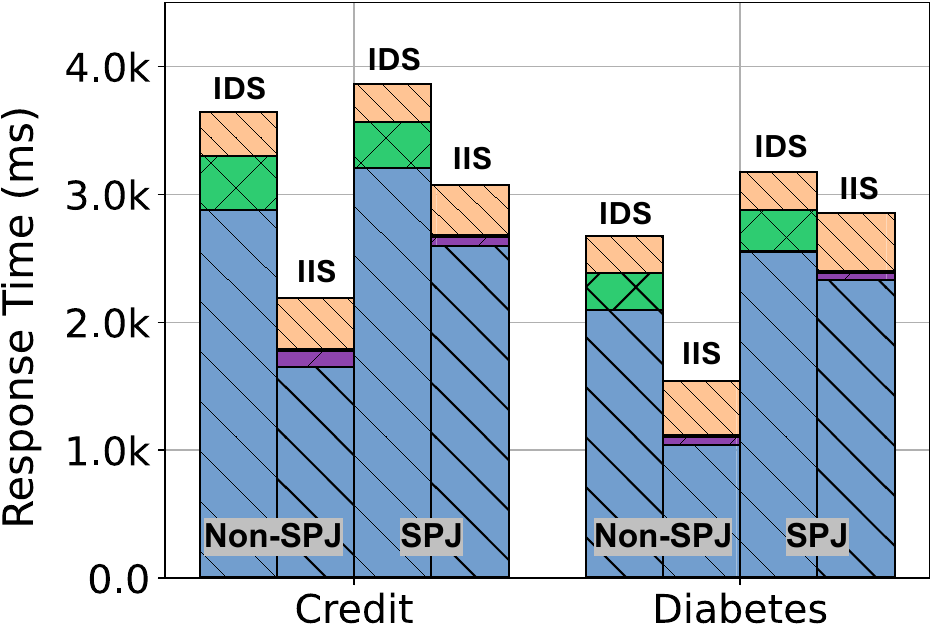}
        \caption{\bluerevise{
        Response time comparison of SPJ and Non-SPJ queries.
        }}
    \label{fig:spj}
\end{subfigure}
\hspace{0.1mm}
\begin{subfigure}[b]{0.23\textwidth}
    \centering
    \includegraphics[width=0.93\columnwidth]{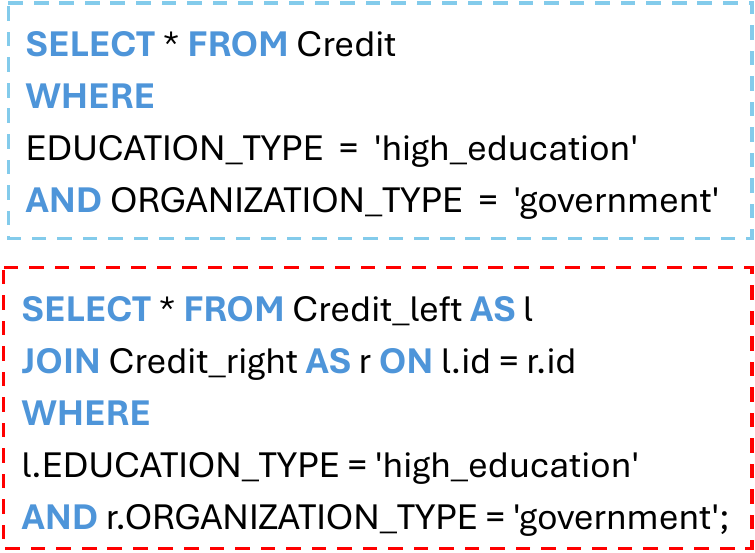}
    \caption{\bluerevise{Example of SPJ (red box) and Non-SPJ queries (blue box).}}
    \label{fig:spj_example}
\end{subfigure}
\caption{\red{Efficiency evaluation of In-database Inference Strategy (IIS) and Inference-Decouple Strategy (IDS).}}
\label{fig:db_exp}
\end{figure*}

\begin{figure}[t]
\centering
\begin{subfigure}[b]{0.45\textwidth}
    \hspace{1mm}
    \centering
\includegraphics[width=0.82\columnwidth]{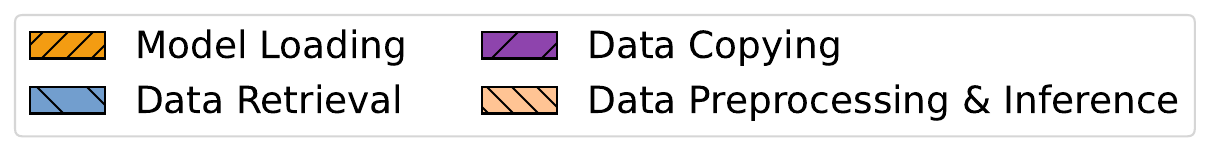}
\end{subfigure}
\vspace{1mm}

\begin{subfigure}[b]{0.49\textwidth}
    \centering
    \includegraphics[width=0.97\columnwidth]{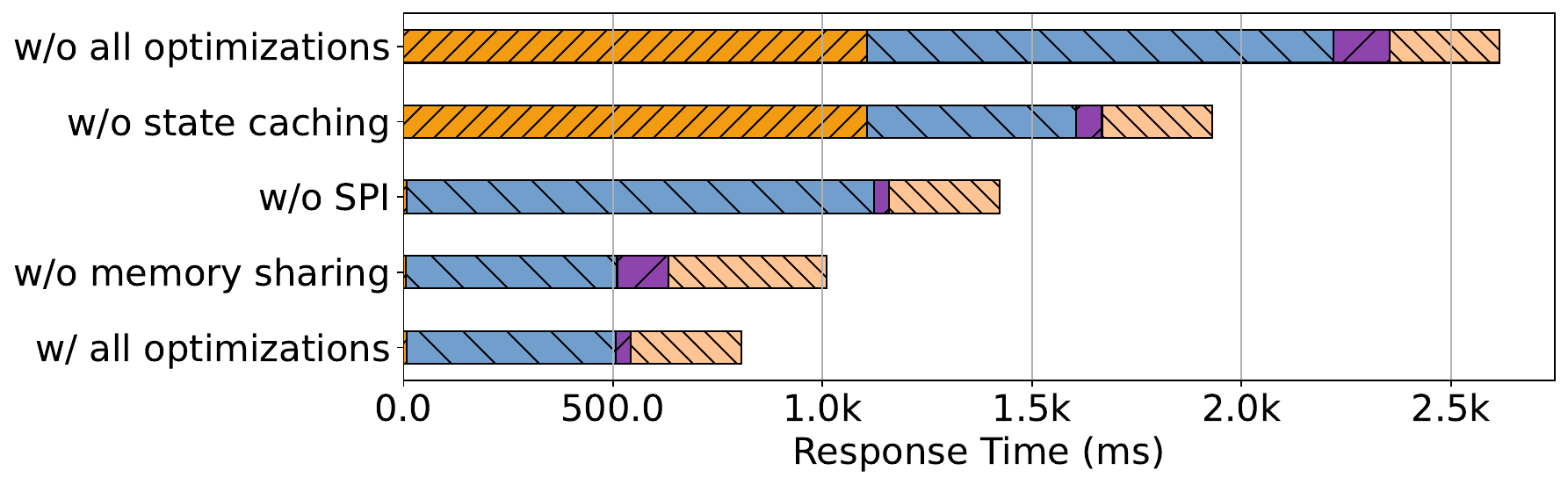}
\end{subfigure}
\caption{Effects of optimizations on response time.}
\label{fig:db_micro}
\end{figure}

\vspace{1mm}
\noindent \textbf{Regularization terms}.
We compare the performance of \Name with three variants:  without the balance term (\textit{\Name w/o $L_{baln}$}), without the sparsity term (\textit{\Name w/o $L_{sprs}$}), and without both terms (\textit{\Name w/o both}). Using DNN as the base model, we conduct experiments on the Payment and Credit datasets.
Figure~\ref{fig:exp_reg_auc} presents the comparison results in w.r.t. Workload-AUC, and Figure~\ref{fig:exp_reg_expert_frequency} analyzes the frequency of expert usage during query execution.
Three main findings emerge: 
First, removing the balance term significantly reduces the Workload-AUC of \Name, Without the balance term, fewer experts are used for each query, leading to lower prediction accuracy. This is evident in Figure~\ref{fig:exp_reg_expert_frequency}, where only two experts are predominately selected.
Second, adding solely the balance term results in lower performance than \Name but utilizes almost all experts for every query. The balance term encourages even expert selection, leading to higher computational costs.
Lastly, enabling both terms simultaneously in \Name balances expert usage while achieving the best performance.

\subsection{System Efficiency}
\label{subsec:RQ3}
\begin{figure}
\centering
\includegraphics[width=0.99\columnwidth]{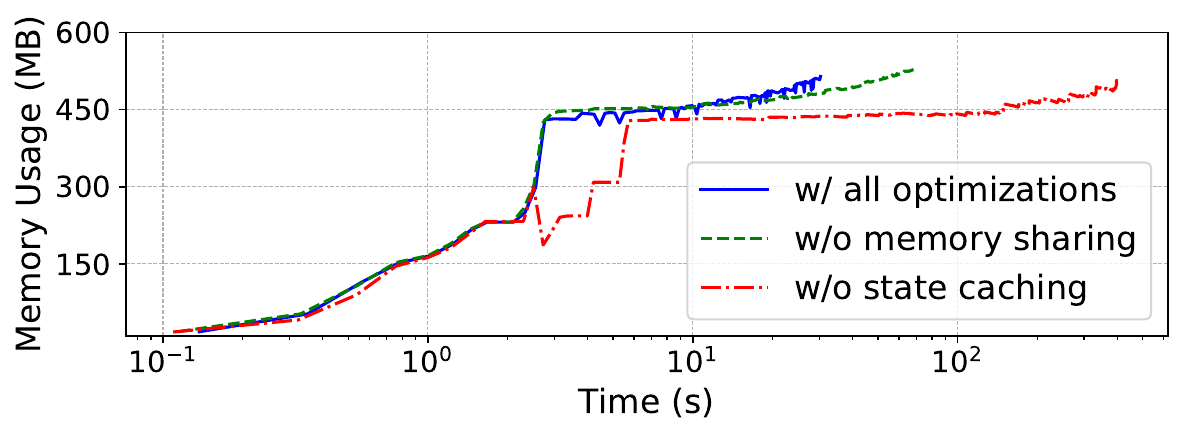} 
\caption{\bluerevise{Optimizing UDFs: analyzing the trade-offs between response time and memory usage.}}
\label{fig:udf_memory}
\end{figure}

\red{To answer question RQ3, we assess the system-level efficiency in end-to-end response time by comparing the IIS that is used in our system with IDS.
}
We implement IIS
as an extension that can be installed in PostgreSQL 14.
\vspace{1mm}
\\\noindent\highlight{Comparison with the baseline.} 
Given an SQL query that selects 100k records for inference,
We report the response time of IIS and IDS on the five datasets, as presented in Figure~\ref{fig:db_macro}. 
Compared to IDS, IIS achieves a speedup of 1.94x, 2.06x, 2.00x, 1.82x, and 1.53x on the Census, Credit, Diabetes, Payment, and Avazu.
There are two reasons for such superior performance. 
One, \red{IIS} reduces the costly data movement overheads between PostgreSQL and the inference system, with lower data retrieval time usage. 
Secondly, \red{IIS} is further enhanced with the optimizations: shared memory to reduce data copying overhead, and 
state caching to eliminate the cost of model loading during the inference UDF execution process. 
\vspace{1mm}
\\\noindent\highlight{Effects of the number of predicting records.} \red{
Next, we examine how the number of predictive records in the SQL query impacts response time. In the Payment dataset, this number ranges from 40k to 640k records.
Figure~\ref{fig:db_scalability} shows the response time for two strategies, IIS and IDS. We observe that IIS consistently surpasses IDS across various record numbers, with performance improvement ranging from 1.47x to 1.93x. 
Moreover, the response time of IIS increases more slowly than that of IDS, since the data movement overhead between the database and the inference system becomes more pronounced with the increase of records. Therefore, with more predictive records, IIS performs even more favorably. }
\vspace{1mm}
\\\noindent
\highlight{Evaluation of optimization techniques} 
Further, we evaluate the benefits of the optimizations mentioned in Section~\ref{subsec:indb_opt}. Specifically, we compare the in-database inference process  with
(i) \textit{w/o memory sharing}; 
(ii) \textit{w/o SPI}; 
(iii) \textit{w/o state caching}; 
and (iv) \textit{w/o all optimizations}. 
Figure~\ref{fig:db_micro} presents the comparison results w.r.t. the response time in predicting 100k records on the Payment dataset. 
The absence of shared memory leads to significant overhead due to data copying between different execution environments. 
Likewise, without SPI (PostgreSQL's built-in data access API), data retrieval times are considerably longer.  Moreover, there is considerable overhead from repeatedly loading the model without state caching. When all optimizations are enabled, the in-database inference extension achieves a 3x speed improvement compared to it without optimizations.
Additionally, we investigate the trade-offs between response time and memory usage brought by memory sharing and state caching. In this analysis, we set a workload containing 300 inference queries and record memory usage during the process. The results presented in Figure~\ref{fig:udf_memory}
reveal that state caching and memory sharing enhance inference efficiency, with minimal impact on overall memory usage from these optimizations.


%

\subsection{Complex Inference Scenarios}
\label{subsec:RQ4}
\bluerevise{
In this subsection, we address question RQ4 by examining \Name's effectiveness under data updates and schema changes, as well as the efficiency of our in-database inference extension when handling SPJ (Select-Project-Join) queries.
}

\begin{figure}[t]
\centering
\captionsetup{justification=centering}
\begin{subfigure}[b]{0.5\textwidth}
    \hspace{1mm}
    \centering
    \includegraphics[width=0.45\columnwidth]{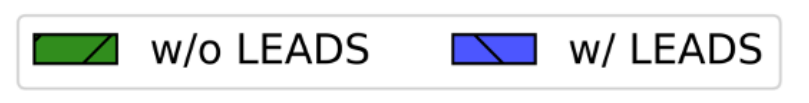}
\end{subfigure}

\begin{subfigure}[b]{0.235\textwidth}
    \centering
    \captionsetup{justification=centering,margin={0cm,0.7cm}}
    \includegraphics[width=0.99\columnwidth]{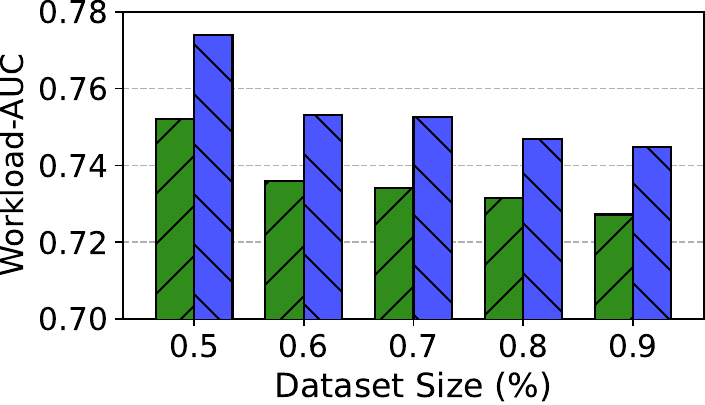}
        \caption{Credit.}
\end{subfigure}
~
\begin{subfigure}[b]{0.235\textwidth}
    \centering
    \captionsetup{justification=centering,,margin={0cm,0.7cm}}
    \includegraphics[width=0.99\columnwidth]{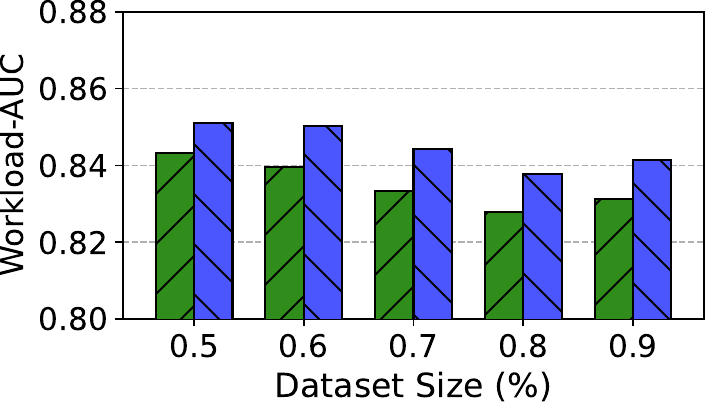}
    \caption{Diabetes.}
\end{subfigure}
\caption{Effects of data insertion on Workload-AUC.}
\label{fig:insertion_workload_auc}
\end{figure}

\vspace{1mm}
\noindent
\textbf{Effectiveness when data updates}. 
We evaluate the effectiveness of \Name for tuple insertion and deletion by executing the same SQL queries between data changes. Using the Credit and Diabetes datasets with a DNN as the base model, each dataset is equally divided into original and new tuples.
The new tuples are inserted into the database in five stages, with the model evaluated after each insertion. As shown in Figure~\ref{fig:insertion_workload_auc}, the model enhanced by \Name consistently outperforms the base model as the dynamic combination of multiple experts discussed in Section~\ref{subsec:RQ1}. In conclusion, when data updates follow the same distribution, the model enhanced by \Name effectively generalizes to various subdatasets.

\begin{figure}[t]
\centering
\begin{subfigure}[b]{0.5\textwidth}
    \hspace{1mm}
    \centering
    \includegraphics[width=0.45\columnwidth]{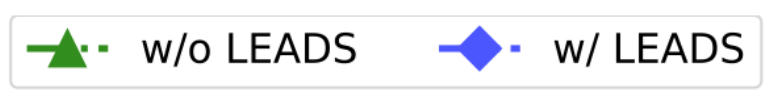}
\end{subfigure}

\begin{subfigure}[b]{0.23\textwidth}
    \centering
    \captionsetup{justification=centering,margin={0cm,0.7cm}}
    \includegraphics[width=0.99\columnwidth]{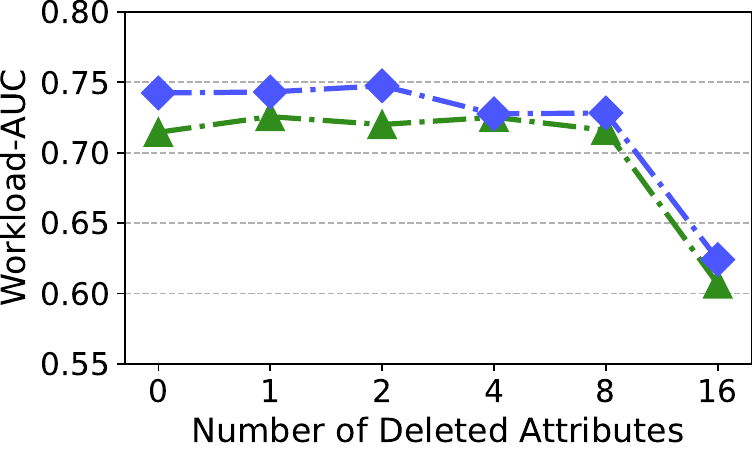}
        \caption{Credit.}
\end{subfigure}
~
\begin{subfigure}[b]{0.23\textwidth}
    \centering
    \captionsetup{justification=centering,,margin={0cm,0.7cm}}
    \includegraphics[width=0.99\columnwidth]{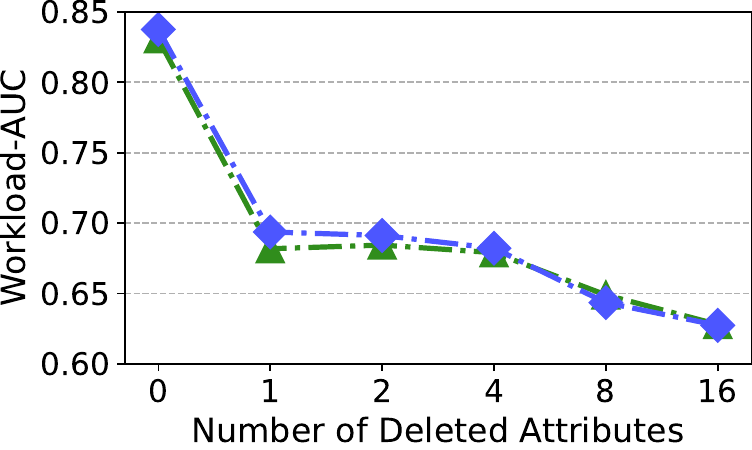}
    \caption{Diabetes.}
\end{subfigure}
\caption{\bluerevise{Effects of schema deletion on Workload-AUC.}}
\label{fig:schema_change}
\end{figure}

\vspace{1mm}
\noindent
\textbf{Effectiveness when the schema changes.}
As for schema changes, altering attributes changes the feature dimension of each tuple, leading to a dimension mismatch with the predictive model. To continue using the original model, we need to construct matched inputs. For insertion, required attributes can be fetched and fed to the model, while newly added attributes will not be considered in predictive modeling. For deletion, the values in the deleted attributes will be missing. We pad these missing values, but this will degrade performance. The impact of deleted attributes on model performance is shown in Figure~\ref{fig:schema_change}.
We notice that as more attributes are deleted, model performance drops rapidly.
Therefore, retraining the model based on the new feature dimensions is advisable to ensure effective inference following schema changes.

\vspace{1mm}
\noindent
\textbf{Efficiency for SPJ queries}. 
\red{
To evaluate in-database inference on SPJ queries, we vertically split the  Diabetes and Credit datasets into two sub-tables (e.g., diabetes\_left and diabetes\_right).
During inference, these sub-tables are joined to select tuples, as shown in the SQL example in  Figure~\ref{fig:spj_example}, and the efficiency is compared to the baseline approach IDS.
We also assess performance using the original table without JOIN to measure the impact of query complexity.
As expected, Figure~\ref{fig:spj} shows that JOIN operations introduce additional overhead, leading to higher response times compared to queries without JOIN.
However, IDS outperforms the baselines due to reduced data movement, as data remains within the database. For JOIN queries, the speed advantage of in-database inference is less pronounced since query planning and execution times dominate over the benefit of eliminating network latency.
}

\section{Related Work}
\label{sec:relatedWork}
\noindent \textbf{Mixture-of-Experts} (MoE)
is initially proposed by \cite{jordan1994hierarchical}  to handle different samples using independent expert modules. \cite{shazeer2017outrageously} introduces the Sparse Gated MoE into language model training, developing large-scale LSTM-based MoE models, in which only one expert is chosen for each input data. 
With the rise of the Transformer as the dominant NLP architecture, researchers began to incorporate MoE layers by extending the Feed-Forward Networks (FFNs) in Transformers to build MoE language models. Despite these innovations, the Sparse Gated MoE \cite{shazeer2017outrageously} struggled with stability due to the expert route strategy where it brute-forcely selects the top-1 expert. To mitigate this, various studies \cite{lepikhin2020gshard, puigcerver2023sparse, fedus2022switch, zhou2022mixture} explored different learnable routing strategies for managing experts and input tokens.
For example,
Fast MoE~\cite{he2021fastmoe} monitors the training status and dynamically adjusts the load for each expert.
Expert Choice Routing~\cite{zhou2022mixture} lets experts choose tokens rather than selecting experts to prevent under-training. 
Soft MoE~\cite{puigcerver2023sparse} performs an implicit soft assignment by passing different weighted combinations of all input tokens to each expert.
In addition to these works on MoE architectures and training strategies, recent years have witnessed the application of MoE in vision-related and multimodal predictive tasks~\cite{riquelme2021scaling}.
However, our research delves into the potential of MoE in structured data analytics.
We closely combine it with database data analytics, dynamically selecting necessary experts corresponding to the filter conditions in SQL queries.

\vspace{1mm}
\noindent \textbf{In-Database Machine Learning} 
involves executing machine learning within the database.
MADlib \cite{hellerstein2012madlib} is an open-source library providing SQL-based ML functions in PostgreSQL. 
Google ML library\cite{GoogleML}, and Microsoft's SQL Server Machine Learning Services \cite{microsoft} offer SQL APIs for ML functions on Oracle, bigquery, and Microsoft SQL Server, respectively.
\bluerevise{
However, they only support feeding the static model with tuples in a data table and do not adapt feeding it with meta-information such as query encoding to customize the model for inference. 
Adding this support requires significant changes to their infrastructure, which is either non-trivial or unfeasible since infrastructure code is not accessible.
Therefore, none of them are directly comparable.
\red{
Complementary to \Name,
 we  propose an effective in-database model selection technique, TRAILS, in ~\cite{xingNAS},
 and we will report their integration in the future.
 }
}

\section{Conclusions}
\label{sec:conclusion}

In this paper, we propose a novel SQL-aware dynamic model slicing technique called \Name. We enhance the general model with the Mixture of Experts (MoE) technique and devise a SQL-aware gating network to dynamically customize a sliced model given the propositional formula in the user's SQL query.
\red{
We further encapsulate \Name into an in-database inference extension for PostgreSQL.
In the implementation, we incorporate three key optimizations to accelerate the in-database inference process. Extensive experiments on five real-world datasets show that \Name consistently outperforms four baseline models, and the in-database inference extension significantly reduces inference time compared to traditional approaches.
We have integrated \Name into NeurDB, our ongoing implementation of an AI-powered autonomous data system.
}

\red{
\subsection*{Acknowledgement:}
This research work is supported by Singapore Ministry of Education Academic Research Fund Tier 3 under MOE’s official grant number MOE2017-T3-1-007.
Yuncheng Wu’s work is supported by National Key Research and Development Program of China (Grant No.2023YFB4503600).
}



\bibliographystyle{ACM-Reference-Format}
\bibliography{reference}

\end{document}